\documentclass[fleqn,usenatbib]{mnras}

\usepackage{newtxtext,newtxmath}

\usepackage[T1]{fontenc}

\DeclareRobustCommand{\VAN}[3]{#2}
\let\VANthebibliography\thebibliography
\def\thebibliography{\DeclareRobustCommand{\VAN}[3]{##3}\VANthebibliography}

\usepackage[normalem]{ulem}

\usepackage{graphicx, caption, subcaption}	
\usepackage{amsmath}	

\newcommand{\Hannah}[1]{{\color{cyan}}}

\newcommand{\el}[2]{\ensuremath{^{#1}\mathrm{#2}}}

\title[Fluorine in the Milky Way]{Chemical Evolution of Fluorine in the Milky Way}

\author[K. A. Womack et al.]{Kate A. Womack,$^{1}$\thanks{E-mail: k.a.womack-2017@hull.ac.uk} Fiorenzo Vincenzo,$^{1}$ Brad K. Gibson,$^{1}$ Benoit C\^{o}t\'{e},$^{2,3}$ Marco Pignatari,$^{3,1,10}$\newauthor Hannah E. Brinkman,$^{3,8,9}$ Paolo Ventura$^{4,7}$ and Amanda Karakas$^{5,6}$
\\
$^{1}$E.~A. Milne Centre for Astrophysics, University of Hull, Hull, HU6 7RX, UK\\
$^{2}$Department of Physics and Astronomy, University of Victoria, Victoria, BC V8P 5C2, Canada \\ 
$^{3}$Konkoly Observatory, Research Centre for Astronomy and Earth Sciences, H-1121 Budapest, Hungary \\
$^{4}$INAF, Observatory of Rome, Via Frascati 33, 00077 Monte Porzio Catone, Italy \\ 
$^{5}$School of Physics \& Astronomy, Monash University, Clayton 3800, Victoria, Australia \\ 
$^{6}$Centre of Excellence for Astrophysics in Three Dimensions (ASTRO-3D), Melbourne, Victoria, Australia \\
$^{7}$Instituto Nazionale di Fisica Nucleare, section of Perugia, Via A. Pascoli snc, 06123 Perugia, Italy \\
$^{8}$ Institute of Astronomy, KU Leuven, Celestijnenlaan 200D, 3001, Leuven, Belgium \\
$^{9}$ Graduate School of Physics, University of Szeged, Dom t\'er 9, Szeged, 6720 Hungary\\
$^{10}$ NuGrid Collaboration, \url{http://nugridstars.org}
}

\date{Accepted XXX. Received YYY; in original form ZZZ}

\pubyear{2022}

\begin{document}
\label{firstpage}
\pagerange{\pageref{firstpage}--\pageref{lastpage}}
\maketitle

\begin{abstract}
 Fluorine has many different potential sites and channels of production, making narrowing down a dominant site of fluorine production particularly challenging. 
In this work, we investigate which sources are the dominant contributors to the galactic fluorine by comparing chemical evolution models to observations of fluorine abundances in Milky Way stars covering a metallicity range -2\,$<$\,[Fe/H]\,$<$\,0.4 and upper limits in the range -3.4\,$<$\,[Fe/H]\,$<$\,-2.3. In our models, we use a variety of stellar yield sets in order to explore the impact of varying both AGB and massive star yields on the chemical evolution of fluorine. In particular, we investigate different prescriptions for initial rotational velocity in massive stars as well as a metallicity dependent mix of rotational velocities. We find that the observed [F/O] and [F/Fe] abundance ratios at low metallicity and the increasing trend of [F/Ba] at [Fe/H]\,$\gtrsim$\,-1 can only be reproduced by chemical evolution models assuming, at all metallicities, a contribution from rapidly rotating massive stars with initial rotational velocities as high as 300\,km s$^{-1}$. A mix of rotational velocities may provide a more physical solution than the sole use of massive stars with $v_{\text{rot}}$\,=\,300\,$\text{km s}^{-1}$, which are predicted to overestimate the fluorine and average s-process elemental abundances at [Fe/H]\,$\gtrsim$\,-1. The contribution from AGB stars is predicted to start at [Fe/H]\,$\approx$\,-1 and becomes increasingly important at high metallicity, being strictly coupled to the evolution of the nitrogen abundance.
Finally, by using modern yield sets, we investigate the fluorine abundances of Wolf-Rayet winds, ruling them out as dominant contributors to the galactic fluorine. 

\end{abstract}




\section{Introduction}\label{sec:intro}
For many years, understanding the origin and evolution of fluorine has posed a challenge for the scientific community. Fluorine has just one stable isotope, \el{19}{F}, with many different channels of production depending on the conditions in stars. \el{19}{F} is also fragile and can be easily destroyed by $\alpha$ captures (e.g. \citealt{Meynet2000}). This makes narrowing down a dominant site for fluorine production particularly difficult. There are five main sites that frequently appear in the literature as having the potential to contribute significantly to the chemical evolution of fluorine; these are the following.
\begin{enumerate}
    \item \textbf{Asymptotic Giant Branch (AGB) stars:} Fluorine is produced in AGB stars during thermal pulses \citep{Forestini1992, Straniero2006}. Secondary \el{19}{F} is made from \el{14}{N} seed nuclei via the following two chains of reactions: \el{14}{N}\,(n,p)\,\el{14}{C}($\rm{\alpha}, \rm{\gamma}$)\,\el{18}{O}\,(p,$\rm{\alpha}$)\,\el{15}{N}\,($\rm{\alpha}, \rm{\gamma}$)\el{19}{F} and \el{14}{N}($\alpha$,$\gamma$)\el{18}{F}($\beta^{+}$)\el{18}{O}(p,$\alpha$)\el{15}{N}($\alpha$, $\gamma$)\el{19}{F} \citep{Lugaro2004}. In certain conditions, primary fluorine can also be made in AGB stars from the rapid burning of \el{13}{C} at high temperatures, which produces \el{15}{N} and allows for the nucleosynthesis of fluorine via \el{15}{N}\,($\rm{\alpha}, \rm{\gamma}$)\el{19}{F} \citep{Cristallo2014}. For a more detailed review of the \el{19}{F} production channels in AGB stars, see \citet{Lucatello2011}. \citet{Kobayashi2011} found that the dominant AGB mass range for fluorine production is $2$-$4\,\text{M}_{\sun}$, over which temperatures do not get hot enough for hot bottom burning to occur, preventing the destruction of fluorine via \el{19}{F}\,($\alpha$, p)\,\el{22}{Ne}. However, it should be noted that the yield set used in \citet{Kobayashi2011} favours \el{19}{F} production in this mass range (see fig. 8 of \citealt{Karakas2007}). There is observational evidence that AGB stars contribute to the galactic fluorine (see the pioneering works of \citealt{Jorissen1992}). However, it is still unclear if AGB stars can account for the total galactic abundance of fluorine.
    \item \textbf{Wolf-Rayet (WR) stars:} Fluorine can be produced by WR stars during the helium burning phase. Again, the seed nuclei for \el{19}{F} production in these stars is \el{14}{N}. If the \el{14}{N} is of secondary origin, then the behavior of \el{19}{F} is also secondary and is, therefore, metallicity dependent. It is thought that WR winds can eject some of the fluorine before it is destroyed by $\alpha$ captures; this process is the result of a delicate balance between the rate at which mass is lost via winds and the efficiency of the \el{19}{F}\,($\alpha$, p)\,\el{22}{Ne} reaction. In one of their models, \citet{Meynet2000} predicted that WR stars can produce as much as $2\times 10^{-3}\,\text{M}_{\rm{\sun}}$ of \el{19}{F}. 
    However, since then other studies have revealed that the \el{19}{F} yield in massive star winds may not be as high as this (e.g. \citealt{Stancliffe2005}, \citealt{Palacios2005}, \citealt{Brinkman2022}). For example, when rotation is accounted for, \citet{Palacios2005} found that the WR fluorine yield falls significantly with respect to \citet{Meynet2000}. Interestingly, \citet{Brinkman2022} found negative net yields of \el{19}{F} in all their rotating and non-rotating models, with the exception of an 80\,M$_{\rm{\sun}}$ model with an initial rotational velocity $v_{\text{rot}}=150\,\text{km s}^{-1}$. All this raises the question - do WR stars contribute to the galactic fluorine budget at all, which will be addressed in later sections of this work.
    \item \textbf{Rotating Massive stars:} Fluorine can be produced in massive stars in the He convective shell via the series of reactions \el{14}{N}\,($\alpha$, $\gamma$)\,\el{18}{F}\,($\beta^{+}$)\,\el{18}{O}\,(p, $\alpha$)\,\el{15}{N}\,($\alpha$, $\gamma$)\,\el{19}{F} (\citealt{Goriely1989}, \citealt{Choplin2018}). This chain of reactions becomes enhanced when rotation is induced, due to the increased abundance of CNO elements which arises as a result of rotation \citep{Limongi2018}. 
    \item \textbf{Core-collapse supernovae (CCSNe):} The $\nu$-process in CCSNe is also a proposed site for fluorine production (\citealt{Woosley1988}, \citealt{Kobayashi2011a}). CCSNe are powered by neutrino heating mechanisms. These neutrinos can interact with some nuclides, including fluorine. \el{19}{F} is produced via the $\nu$ process in CCSNe by the following reaction: \el{20}{Ne}\,($\nu$, $\nu^{'}$\,p)\el{19}{F}. Exactly how much fluorine this process might produce in CCSN is unclear because there is uncertainty around the flux and energy of the neutrinos. However, given this production is a primary process, more observations at low metallicity might help us to constrain how much fluorine we might expect to be produced by this source.
    \item \textbf{Novae:} \citet{Jose1998} showed that fluorine can be produced by novae. The mechanism for novae to produce fluorine is as follows: \el{17}{O}\,(p, $\gamma$)\,\el{18}{F}\,(p, $\gamma$)\,\el{19}{Ne}\,($\beta^{+}$)\,\el{19}{F}. Just as with the $\nu$-process, fluorine yields from novae are still highly uncertain. \citet{Jose1998} found that fluorine was only significantly synthesised in their $1.35\,\text{M}_{\sun}$ models. Therefore, we cannot be sure of their contribution to the galactic fluorine abundance.
\end{enumerate}
Note that here and throughout this work we define AGB stars in the mass range 1\,$\leq$\,M/M$_{\rm{\odot}}$\,$\leq$\,8 and massive stars 8\,$<$\,M/M$_{\rm{\odot}}$\,$\leq$\,120. 

Many chemical evolution studies have tried to disentangle this web and figure out which sources of fluorine are dominant in different metallicity ranges. There is not much agreement between authors. \citet{Renda2004} used the WR yields of \citet{Meynet2000} to show that WR stars can dominate fluorine production at solar and super-solar metallicities, while AGB stars were required in their models to reproduce the trends at lower metallicities. This is in contrast to the work of \citet{Olive2019} who concluded that AGB stars dominate at high metallicity and that the $\nu$-process in CCSN is required to reproduce low-metallicity observations. A combination of AGB stars and neutrino process was also used by \citet{Kobayashi2011a} to reproduce the observed behaviour of [F/O] in globular cluster and solar neighbourhood stars. 

\citet{Timmes1995} was the first chemical evolution study to investigate fluorine, and they found that the inclusion of novae can reproduce [F/O] ratios in combination with AGB stars. The need for novae to reproduce [F/O] versus [O/H] ratios was also found by \citet{Spitoni2018}, who concluded that AGB and WR stars dominate galactic fluorine production. We note again that \citet{Kobayashi2011} found that the dominant AGB mass range that contributes to fluorine is $2$-$4\,\text{M}_{\sun}$ but this contribution can only be seen at $[\text{Fe/H}]\gtrsim -1.5\,\text{dex}$.

By assuming that massive stars have, on average, increasingly faster initial rotational velocities at low metallicities, \citet{Prantzos2018} found that rotating massive stars can dominate the evolution of fluorine in the Solar Neighbourhood up to solar metallicity. A similar conclusion was reached by \citet{Grisoni2020}, who investigated the chemical evolution of fluorine by separately modelling the thick and the thin disk of the Milky Way using the so-called `parallel model' of \citet{Grisoni2017}. In particular, \citet{Grisoni2020} concluded that rotating massive stars can dominate fluorine production up to solar metallicity but a boost in fluorine is also needed at higher metallicities in order to match the behaviour of the observations. They proposed that this boost could be obtained either by artificially enhancing the AGB yields or by including an additional contribution from novae in the models. The prescription for rotating massive stars in \citet{Grisoni2020} follows the assumptions of \citet{Romano2019} where all stars with $[\text{Fe/H}]$\,<\,-1\,dex are given an initial rotational velocity $v_{\text{rot}}$\,=\,300\,$\text{km s}^{-1}$ while all stars with [Fe/H]\,$\ge$\,-1\,dex have $v_{\text{rot}}$\,= 0\,$\text{km s}^{-1}$. Rotating massive stars were first recognised as important at low metallicity by \citet{Chiappini2006} in relation to primary nitrogen production, which is the seed for fluorine production. This arose from the work of \citet{Matteucci1986} who recognised the need for another primary component of nitrogen.

Fluorine has also recently become an element of interest  for high redshift studies. \citet{Franco2021} were able to estimate the abundance of fluorine  in a gravitationally lensed galaxy at a redshift of z\,=\,4.4, determining that Wolf-Rayet stars must be responsible for the observed fluorine abundance enhancement. Though this is not a Milky Way observation, it can still give us an idea of the origins of fluorine in the early Universe and thus, presumably, at low metallicity.

Aside from the uncertainties in the dominant production site of fluorine, we must also contend with difficulty in gathering observations of fluorine. The majority of fluorine abundance determinations in the literature are obtained from the analysis of ro-vibrational HF lines at 2.3\,$\mu$\,m \citep{Abia2015}. This spectral range is contaminated by lots of telluric lines, which prevent the use of many HF lines for fluorine abundance determinations. Recently, the first detection of an AlF line was obtained in 2 M-type AGB stars \citep{Saberi2022}. \citet{Danilovich2021} also detected the AlF line towards an S-type AGB star, measuring an abundance of AlF 40\% greater than solar. 

Most fluorine observations for chemical evolution studies are available using HF lines as detected in both galactic and extra-galactic AGB stars \citep{Abia2011, Abia2015, Abia2019}, field stars (\citealt{Lucatello2011}, \citealt{Li2013}) and in the Galactic center \citep{Guerco2022}. There are also a variety observations of fluorine in open and globular clusters (e.g. \citealt{Maiorca2014}, \citealt{Nault2013}, \citealt{deLaverny2013}, \citealt{Smith2005}, \citealt{Cunha2003, Cunha2005}, \citealt{Yong2008}). Since this work is mainly focused on the chemical evolution of fluorine in Milky Way field stars, the previously listed observations in open and globular clusters will not be included in our analysis. 

The evolution of fluorine at low metallicity (e.g., [Fe/H]\,$\lesssim$\,-1.5\,dex) poses a particular challenge because of a large contamination from telluric lines and blending of the HF lines with CO features \citep{Lucatello2011}. Despite those challenges, there are some measurements of fluorine abundances at low metallicities,  which include a sample of red giants from \citet{Lucatello2011} and two red giants in Carina dwarf spheroidal (dSph) galaxy from \citet{Abia2015} among others (e.g. \citealt{Li2013}, \citealt{MuraGuzman2020}). Both the stellar sample of \citet{Lucatello2011} and the Carina stars from \citet{Abia2015} are considered in our work. 

The structure of the paper is as follows: Section \ref{sec:obs} lays out the sample of fluorine abundance measurements that are used in this work for different metallicity ranges, Section \ref{sec:model} introduces the main hypothesis and working assumptions of our galactic chemical evolution model and summarizes the different combinations of yields that are included in the model, Section \ref{sec:results} presents the main chemical evolution trends of interest as predicted by our model to reproduce observational data, and Section \ref{sec:discussion} explains how these results can help us to probe the chemical evolution of fluorine. Finally, in Section \ref{sec:conclusions}, we present our conclusions.



\section{Observations}\label{sec:obs}
The most recent set of fluorine abundance measurements are those of \citet{Ryde2020} who observed 66 red giants using the Immersion GRating INfrared Spectrometer (\textit{IGRINS}) and the \textit{Phoenix} infrared high-resolution spectrograph at the \textit{Gemini} South Observatory. The metallicity range of these observations is -1.1\,<\,[Fe/H]\,<\,0.4 which extends the metallicity range of fluorine abundances in the solar neighbourhood that were available previous to this study (e.g. \citealt{Jonsson2017}).

Due to telluric lines and blending, much of the data we have at low [Fe/H] are upper limits rather than absolute measurements. Though not as conclusive as absolute measurements, upper limits can still tell us about the range of fluorine abundances we might expect and can give us a preliminary idea of if our chemical evolution models can reproduce observations at low metallicity. The primary set of fluorine observations at low metallicity used in this work consists of a sample of eleven metal-poor red giant stars from \citet{Lucatello2011}. The abundances were measured from the analysis of spectra obtained with the CRyogenic high-resolution InfraRed Echelle Spectrograph (\textit{CRIRES}) on ESO's \textit{VLT}. Of the 11 stars in the metallicity range -3.4\,<\,[Fe/H]\,<\,-1.3, two have abundance measurements of fluorine, while the remaining nine have upper limits provided. 

Eight red giants in the sample of \citet{Lucatello2011} are classified as CEMP-s stars (carbon-enhanced metal poor stars that are also enriched in s-process elements), whereas two stars are classified as CEMP-no star (not enriched with s-process or r-process elements). There is also one star in this sample classified as carbon normal. While the physical origin of CEMP-no stars is still unclear and debated \citep{Aoki2002,Yoon2016,Hansen2016}, the s-process and carbon enhancement as measured in the atmosphere of CEMP-s red giants likely results from binary mass transfer from an AGB companion that changed the initial surface abundances (e.g., \citealt{Lucatello2005,Beers2005,Bisterzo2010,Lugaro2012,Starkenburg2014,Hansen2016a,Hampel2016}). Therefore, the predictions of our chemical evolution models at low [Fe/H] solely provide a baseline for the average fluorine abundances at birth in CEMP-s red giants before mass transfer took place. 
We also include fluorine measurements as obtained in two stars of the Carina dSph galaxy by \citet{Abia2015}. These measurements were obtained from spectra taken using the \textit{Phoenix} infrared high-resolution spectrograph by \citet{Abia2011} and reanalysed by using the spectral synthesis code {\tt Terbospectrum} by \citet{Abia2015}. The formation of Carina occurred with low star formation efficiencies and a short infall timescale (e.g. \citealt{Lanfranchi2006}, \citealt{Vincenzo2014}), as did the Milky Way halo. Therefore, observations in Carina dSph have been included in this work in order to further our understanding of how fluorine might behave at low metallicity in general. However, since these stars are not Milky Way stars we must be careful as they are not directly comparable with the chemical evolution models presented in this work or the other observations. The chemical evolution of fluorine in Carina will be the subject of future work.

\section{Galactic Chemical Evolution Model}\label{sec:model}
We have used the chemical evolution code {\tt OMEGA+}\footnote{{\tt OMEGA+} is available online as part of the JINAPyCEE package \url{https://github.com/becot85/JINAPyCEE}} \citep{Cote2018}. This is a two-zone model where a central star forming region is modelled using the code {\tt OMEGA}\footnote{{\tt OMEGA} is available online as part of the NuPyCEE package \url{https://github.com/NuGrid/NUPYCEE}}\citep{Cote2017}, which simulates the evolution of several physical and chemical properties within a cold gas reservoir, surrounded by a non-star forming hot gas reservoir. The latter is considered as the circumgalactic medium (CGM) in our model.

We can follow both the evolution of the CGM and the internal star forming galaxy. The evolution of the mass of the gas in the CGM ($M_{\rm CGM}$) is as follows:
\begin{equation}
    \dot{M}_{\rm CGM}(t) = \dot{M}_{\rm CGM,in}(t)  + \dot{M}_{\rm outflow}(t) - \dot{M}_{\rm inflow}(t) - \dot{M}_{\rm CGM,out}(t)
\end{equation}
where $\dot{M}_{\rm CGM,in}$ is the inflow rate from the external intergalactic medium into the CGM, $\dot{M}_{\rm outflow}$ is the mass removed from the central galaxy and added to the CGM via outflow, $\dot{M}_{\rm inflow}$ is the gas which flows into the galaxy from the CGM and $\dot{M}_{\rm CGM,out}$ is the outflow rate of gas from the CGM into the intergalactic medium. The intergalactic medium represents the space outside the CGM and is defined as a sphere with radius equal to the virial radius of the dark matter halo that hosts the central galaxy. The mass of the CGM tends to increase if the mass of the dark matter halo also increases, as $\dot{M}_{\rm CGM,in}$ can reach higher values due to a larger availability of gas in the environment; conversely, the CGM mass will decrease when the mass of the dark matter halo decreases, as gas can more efficiently leave the CGM, giving rise to higher values of $\dot{M}_{\rm CGM,out}$. We can also decrease the mass of the gas in the CGM, even if the dark matter mass stays constant, by allowing the CGM to have large scale outflows. Details of all of these terms can be found in \citet{Cote2018, Cote2019} and references therein (see fig. 7 of \citet{Cote2018} for a visual representation of the workings of {\tt OMEGA+}).

The evolution of the galactic gas mass $\dot{M}_{\rm gas}$ is defined as (\citealt{Tinsley1980}; \citealt{Pagel1997}; \citealt{Matteucci2012}):
\begin{equation}
    \dot{M}_{\rm gas}(t) = \dot{M}_{\rm inflow}(t)  + \dot{M}_{\rm ej}(t) - \dot{M}_\star(t) - \dot{M}_{\rm outflow}(t)
\end{equation}
where $\dot{M}_{\rm inflow}$ is the mass added by galactic inflows from the CGM, $\dot{M}_{\rm ej}$ is the mass added by stellar ejecta, $\dot{M}_\star$ is the mass locked away by star formation and $\dot{M}_{\rm outflow}$ is the mass lost by outflows into the CGM. This equation is used at each timestep to track the evolution of the galaxy across 13\,Gyr.

The infall prescription of gas from the CGM into the galaxy we use here is a dual infall model based on \citet{Chiappini1997}. It combines two episodes of exponential gas inflow and is described as follows:
\begin{equation} \label{eq:infall}
    \dot{M}_\mathrm{inflow}(t)=A_{1}\mathrm{exp}\left(\frac{-t}{\tau_{1}}\right) + A_{2}\mathrm{exp}\left(\frac{t_{\mathrm{max}} - t}{\tau_{2}}\right).
\end{equation}
Where A$_{1}$, A$_{2}$, $\tau_{1}$, $\tau_{2}$, and t$_{\mathrm{max}}$ are free parameters, the values for which can be found in Table \ref{tab:model_properties}. A$_{1}$ and A$_{2}$ represent the normalisation of the first and second infall events, respectively, $\tau_{1}$ and $\tau_{2}$ are the timescales for mass accretion for the first and second infall, and t$_{\mathrm{max}}$ is the time of maximum contribution of the second gas accretion episode, which is zero for the first episode.

The star formation rate is defined as
\begin{equation}\label{eq:sfr}
    \dot{M}_\star(t)\,=\,\frac{\epsilon_\star}{\tau_\star}\,M_{\rm gas}(t)
\end{equation}
where $\epsilon_\star$ and $\tau_\star$ are the dimensionless star formation efficiency (sfe) and star formation timescale, respectively. The outflow rate is proportional to the star formation rate and is defined as
\begin{equation}\label{eq:outflow}
    \dot{M}_{\rm outflow}(t)\,=\,\eta\,\dot{M}_\star(t)
\end{equation}
where $\eta$ is the mass loading factor and controls the strength of the outflows. The values for $\epsilon_\star$, $\tau_\star$ and $\eta$ can also be found in Table \ref{tab:model_properties}.

To calculate the mass of gas added by stellar ejecta, the contribution of every stellar population formed by time $t$ is summed so that
\begin{equation}
    \dot{M}_{\rm ej}(t)\,=\,\sum_{j}\,\dot{M}_{\rm ej}^{j}\,(M_{j}, Z_{j}, t-t_{j})
\end{equation}
where $\dot{M}_{\rm ej}^{j}$ is the mass ejected by the $j$th stellar population, $M_{j}$ is the initial mass of the population, $Z_{j}$ is the initial metallicity of the population and $t-t_{j}$ is the age of the $j$th population at time $t$. The simple stellar populations (SSPs) are created at every timestep using {\tt SYGMA} (Stellar Yields for Galactic Modelling Applications) \citep{Ritter2018}. An SSP is defined as a population of stars with the same age and chemical composition, with the number of each type of star in the different evolutionary stages being weighted by an initial mass function (IMF). In this work we adopt the IMF of \citet{Kroupa2001}. {\tt SYGMA} includes ejecta from low and intermediate mass stars, massive stars, Type Ia Supernovae (SNe Ia), neutron star mergers and additional sources can also be added manually by the user. The ejecta from the SSPs are then instantaneously and uniformly mixed into the gas reservoir. 

SNe Ia are modelled by assuming a power-law delay-time distribution (DTD) similar to that of \citet[see also \citealt{Freundlich2021,Wiseman2021}]{Maoz2012} in the form $t^{-\beta}$ with $\beta$\,=\,1. The minimum delay time of SNe Ia is set by the lifetime of intermediate-mass stars used in the galactic chemical evolution (GCE) calculation. For every SSP, at any time $t$, the DTD is multiplied by the fraction of progenitor white dwarfs ($f_{\text{WD}}(t)$) originating from stars in the mass range of 3 to 8 M$_{\sun}$ (see \citealt{Ritter2018} for more details). $f_{\text{WD}}(t)$ smoothly  evolves from 0 to 1 when the age of the SSP transits from the lifetime of a 8 M$_{\sun}$ star to the lifetime of a 3 M$_{\sun}$ star. The temporal evolution of the rate of SNe Ia is normalized such that 10$^{-3}$ SN occurs per units of solar mass formed (see Table 5 in \citealt{Cote2016} for references).
We use the solar abundances of \citet{Asplund2009} throughout, where the \el{19}{F} solar abundance is $\log\big(\epsilon\,(\text{F})\big) = 4.56 \pm 0.30$.

\subsection{Stellar Yields}\label{sec:params_and_yields}
\begin{table}
    \centering
    \begin{tabular}{l|c}
        Parameter & Value \\
        \hline
        A$_{1}$ [M$_\mathrm{\sun}$\,yr$^{-1}$] & 46 \\
        A$_{2}$ [M$_\mathrm{\sun}$\,yr$^{-1}$]& 5.9 \\
        $\tau_{1}$ [Gyr] &  0.8 \\
        $\tau_{2}$ [Gyr] & 7.0 \\
        t$_{\rm{max}}$ [Gyr] & 1.0\\
        $\epsilon_\star$& 0.23\\
        $\tau_\star$ [Gyr] & 1.0\\
        $\eta$ & 0.52\\
    \end{tabular}
    \caption{Parameter values of the model, where A$_{1}$, A$_{2}$, $\tau_{1}$, $\tau_{2}$ and t$_{\rm{max}}$ are all free parameters of Equation \ref{eq:infall}. $\epsilon_\star$, the sfe, and $\tau_\star$, the star formation timescale, are the free parameters of Equation \ref{eq:sfr} and $\eta$, the mass loading factor, is the free parameter of Equation \ref{eq:outflow}. These values are equivalent to the values in the `best' model of \citet{Cote2019}.}
    \label{tab:model_properties}
\end{table}

Nine combinations of yields have been used throughout this work and are laid out in Table \ref{tab:yields}. We explore the following options for our AGB yields: \textit{(i)} the FUll-Network Repository of Updated Isotopic Tables \& Yields (\textit{FRUITY}) for AGB stars from \citet{Cristallo2015}, that are available for metallicities $10^{-4}\leq Z \leq 2\times 10^{-2}$ and masses in the range $1.3$-$6.0\,\text{M}_{\sun}$; \textit{(ii)} the Monash AGB yields from \citet{Lugaro2012,Karakas2016}, and \citet{Karakas2018} with metallicities $10^{-4}\leq Z \leq 3\times 10^{-2}$ and masses in the range $0.9$-$8.0\,\text{M}_{\sun}$; \textit{(iii)} an extended version of the previous Monash yields that cover the same range of masses and metallicities as the previous set, where heavy elements (anything heavier than iron) are also included \citep{Karakas2016, Karakas2018}; finally, \textit{(iv)} the AGB yields from \citet{Ventura2013, Ventura2014, Ventura2018} with metallicities $3\times 10^{-4}\leq Z \leq 1.4\times 10^{-2}$ and masses in the range $1.0$-$7.5\,\text{M}_{\sun}$.

We consider the two following options for our massive star yields. 
\begin{enumerate}
    \item Set R of \citet{Limongi2018}, who developed stellar evolution models for massive stars by assuming three different initial rotation velocities as follows: $v_{\text{rot}}=300\,\text{km s}^{-1}$, $v_{\text{rot}}=150\,\text{km s}^{-1}$, and no rotation; all of these options will be explored in this work. For each rotational velocity, \citet{Limongi2018} developed models with initial iron abundances $[\text{Fe/H}]=0$, $-1$, $-2$, and $-3\,\text{dex}$ in the mass range $13$-$120\,\text{M}_{\sun}$. The chemical evolution models of \citet{Prantzos2018} assume a yield set which combines the massive star models of \citet{Limongi2018} with different $v_{\text{rot}}$ depending on metallicity, by assuming that lower metallicity stars rotate faster, on average, than higher metallicity stars, as illustrated in fig. 4 of \citet{Prantzos2018}. A similar mixture of rotating massive star models that varies as a function of [Fe/H] will also be explored in this work. The logic for this combination comes about because \citet{Meynet1997} stated that in order to conserve angular momentum, low metallicity stars must rotate faster as they are more compact.
    \item The yields of \citet[see also \citealt{Kobayashi2006,Kobayashi2011a}, and \citealt{Kobayashi2020}]{Nomoto2013} which do not include rotation. These yields use metallicities $10^{-3}\leq Z \leq 5\times 10^{-2}$ in the mass range $13$-$40\,\text{M}_{\sun}$. 
\end{enumerate}

Of these yield sets, we mainly consider the \textit{FRUITY} AGB yields because they cover a large range of masses and metallicities. The code used to calculate these yields is also coupled to a full nuclear network up to the termination point of the s-process, therefore it considers the full range of isotopes and reactions relevant to this work. For massive stars, we mainly use the yields of \citet{Limongi2018} in order to investigate the impact of rotation. Finally, for all models we use the W7 SNIa yields of \citet{Iwamoto1999}.

\begin{table}
    \centering
    \begin{tabular}{c|c|c|c}
         Model Name & AGB yields & Massive Star Yields & SNIa Yields  \\
         \hline
         CLCmix & FRUITY & L\&C V$_{\rm{rot}}$ = mix & Iwamoto \\
         CLC000 & FRUITY & L\&C V$_{\rm{rot}}$ = 0\,kms$^{-1}$ & Iwamoto \\
         CLC150 & FRUITY & L\&C V$_{\rm{rot}}$ = 150\,kms$^{-1}$ & Iwamoto \\
         CLC300 & FRUITY & L\&C V$_{\rm{rot}}$ = 300\,kms$^{-1}$ & Iwamoto \\
         Mon18LCmix & Mon. 1 & L\&C V$_{\rm{rot}}$ = mix & Iwamoto \\
         MonLCmix & Mon. 2 & L\&C V$_{\rm{rot}}$ = mix & Iwamoto \\ 
         MonLC300 & Mon. 2 & L\&C V$_{\rm{rot}}$ = 300\,kms$^{-1}$ & Iwamoto \\
         CNom & FRUITY & Nomoto & Iwamoto \\
         VenLCmix & ATON & L\&C V$_{\rm{rot}}$ = mix & Iwamoto
    \end{tabular}
    \caption{Combination of yields used for the chemical evolution modelling in this work. Where FRUITY = \citet{Cristallo2015}, Mon 1 = \citet{Lugaro2012} and \citet{Karakas2016, Karakas2018}, Mon 2 is the same as previous but with heavy elements included and ATON = \citet{Ventura2013, Ventura2014, Ventura2018}. L\&C are the massive star yields of \citet{Limongi2018}, Nomoto are the yields of \citet*{Nomoto2013} and Iwamoto are the SN1a yields of \citet{Iwamoto1999}.}
    \label{tab:yields}
\end{table}

\section{Results}\label{sec:results}
\begin{figure*}
\centering
\begin{subfigure}[t]{0.49\textwidth}
 \includegraphics[width=\textwidth]{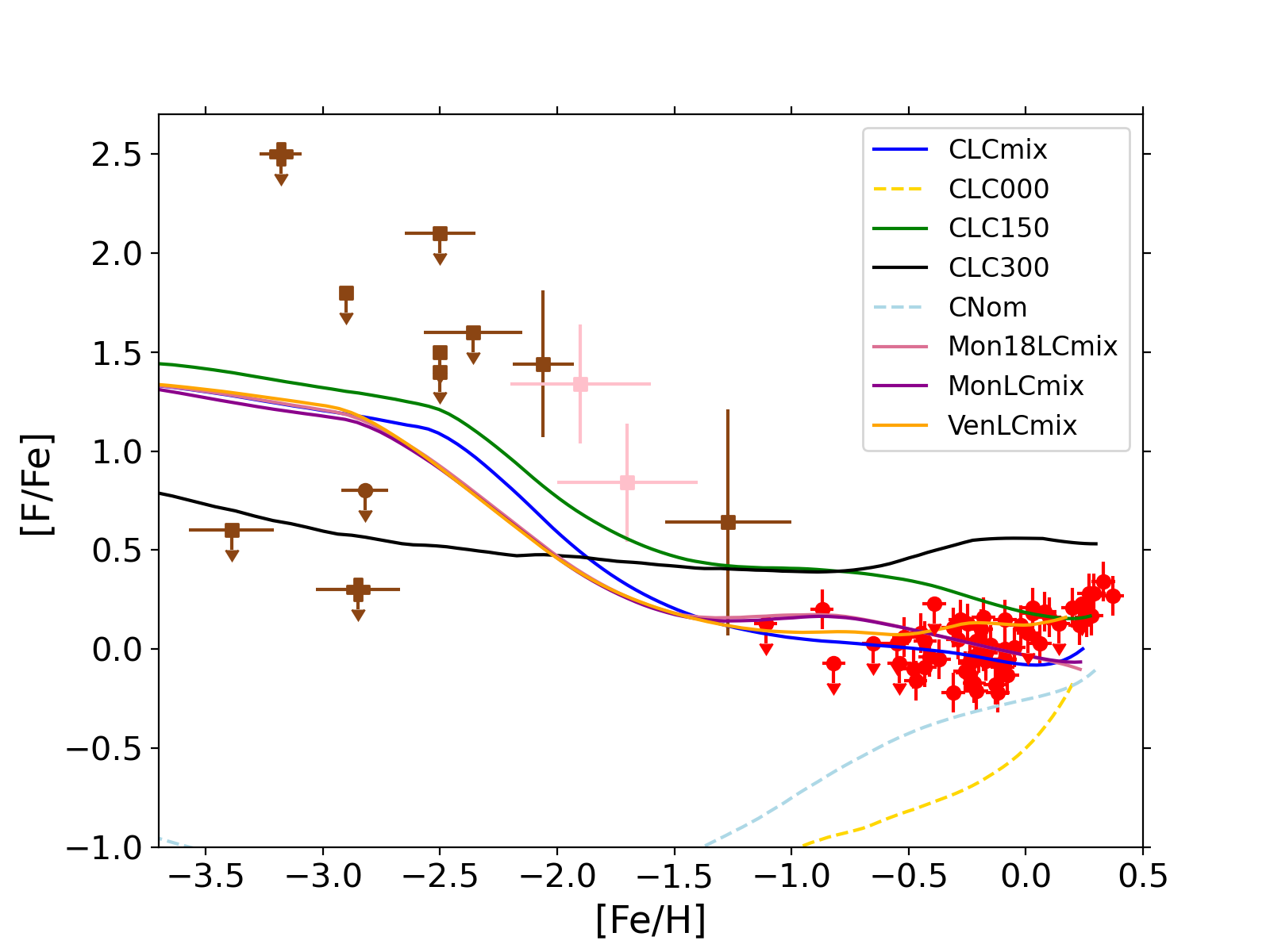}
 \caption{[F/Fe] versus [Fe/H]}
 \label{fig:FFe_FeH}
\end{subfigure}
\hfill
\begin{subfigure}[t]{0.49\textwidth}
 \includegraphics[width=\textwidth]{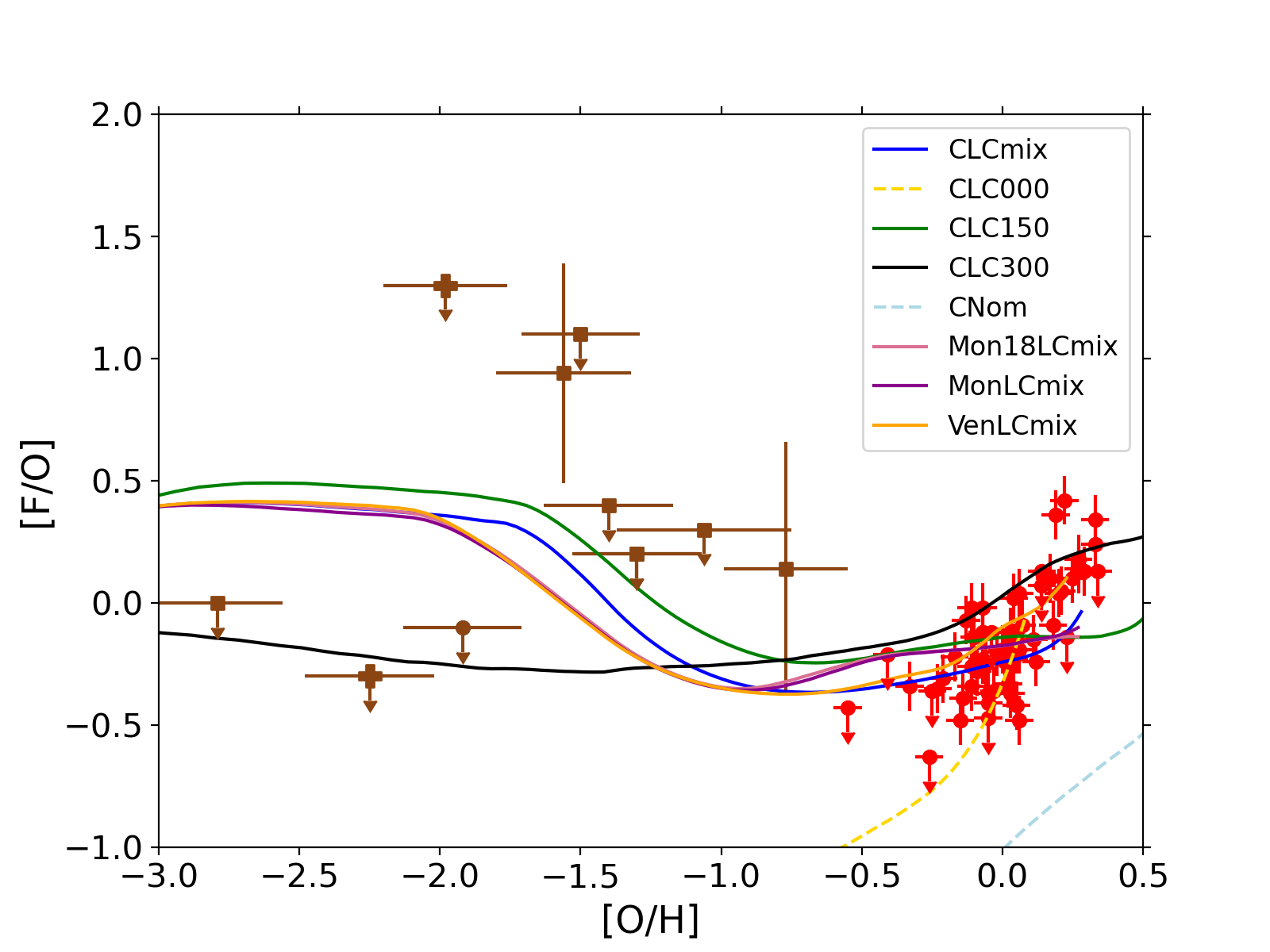}
 \caption{[F/O] versus [O/H]}
 \label{fig:FO_OH}
\end{subfigure}
 \caption{Left-hand panel: [F/Fe] versus [Fe/H] for models CLCmix, CLC000, CLC150, CLC300, CNom, Mon18LCmix and MonLCmix. The red points are observations of fluorine from \citet{Ryde2020}, the brown points from \citet{Lucatello2011} and the pink squares are Carina data from \citet{Abia2015}. CEMP-s stars in the sample of \citet{Lucatello2011} are represented by brown squares, CEMP-no stars by brown crosses and the carbon normal star by a brown point. Right-hand panel: the same but for [F/O] versus [O/H].}
 \label{fig:Comb_FO}
\end{figure*}

Fig.~\ref{fig:FFe_FeH} shows the abundance trend of [F/Fe] versus [Fe/H] for models assuming different combinations of yields, as summarised in Table \ref{tab:yields}. The predictions of our models are compared with high metallicity observations of fluorine abundances in the red giant sample of \citet[red points with error bars]{Ryde2020} and low metallicity observations in red giants from  \citet[brown square symbols represent stars classified as CEMP-s  whereas brown crosses are CEMP-no stars, the carbon normal star is represented by a brown point]{Lucatello2011} and \citet[pink squares with error bars]{Abia2015}. We remind the readers that our models predict the evolution of the chemical abundances in the interstellar medium (ISM), hence how the birth abundances of stars change with time throughout the evolution of the Galaxy. 

The two models which include massive stars with no rotation (\textit{CLC000} and \textit{CNom}) show an increasing trend in [F/Fe] at higher metallicities that is in line with observations. However, these two models lie below the observed abundances. We note that, in the low metallicity regime from -3.5\,$<$\,[Fe/H]\,$<$\,-2, the observational data are upper limits rather than absolute measurements, along with the high dispersion of the observational data in this metallicity range, which prevents us from drawing strong conclusions in this regime. The strongest constraint on our chemical evolution models is provided by observations in the metallicity range -0.7\,$<$\,[Fe/H]\,$<$\,0.4. We can still draw conclusions in the range -2\,$<$\,[Fe/H]\,$<$\,-0.7 but we are limited by poor statistics. All models with rotating massive stars included cut through the middle of the upper limits, with the majority of the brown points sitting above the chemical evolution trend lines. This is important because we know that an upper limit means the value quoted has the potential to be lower than what is measured. We also see that the models including rotating massive stars are consistent with the bulk of the high-metallicity data.
We note that models with higher $v_{\text{rot}}$ can reach increasingly higher $[\text{F/Fe}]$ ratios at high metallicity. Nonetheless, no model with high $v_{\text{rot}}$ reproduces the upward trend seen in observations at high [Fe/H], which is only seen in the models with $v_{\text{rot}}=0$.

Fig.~\ref{fig:FO_OH} shows the abundance trend of [F/O] versus [O/H] for the same set of models as in Fig.~\ref{fig:FFe_FeH}. These ratios are commonly plotted together when studying the chemical evolution of fluorine to trace the impact of the chemical enrichment from massive stars with minimal connection to the choice of the location of the mass cut in the massive star models. Looking at chemical evolution trends relative to oxygen is also useful because they do not include the uncertainties associated with SNIa models.
In Fig.~\ref{fig:FO_OH} the trajectories are again compared with observations from \citet{Ryde2020} and \citet{Lucatello2011}. Again, some observations provide better constraints for our GCE models than others, with those at [O/H]\,$>$\,-0.4 providing the strongest constraint. We can see that all models which include any sort of prescription for rotation in massive stars cut through the low metallicity observations, including \textit{VenLCmix}. Further discussion of this model can be found later in the section. Of the two models that do not include rotating massive stars (\textit{CLC000} and \textit{CNom}), only \textit{CLC000} reproduces the abundance trends of the high metallicity observations.

In Figs \ref{fig:FFe_FeH} and \ref{fig:FO_OH} we also investigate the impact of different AGB stellar yields on the chemical enrichment of fluorine. Our model with the \textit{FRUITY} stellar yields for AGB stars (\textit{CLCmix}) predicts similar abundance trends as the models with the Monash stellar yields (\textit{Mon18LCmix} and \textit{MonLCmix}). The model with the AGB stellar yields of \citet{Ventura2013, Ventura2014, Ventura2018} (\textit{VenLCmix}) predict higher final fluorine abundances in both Figs. \ref{fig:FFe_FeH} and \ref{fig:FO_OH} compared to the \textit{FRUITY} and Monash yields but they still lie within the high metallicity observations.

\begin{figure}
\centering
\begin{subfigure}[t]{0.49\textwidth}
 \includegraphics[width=\textwidth]{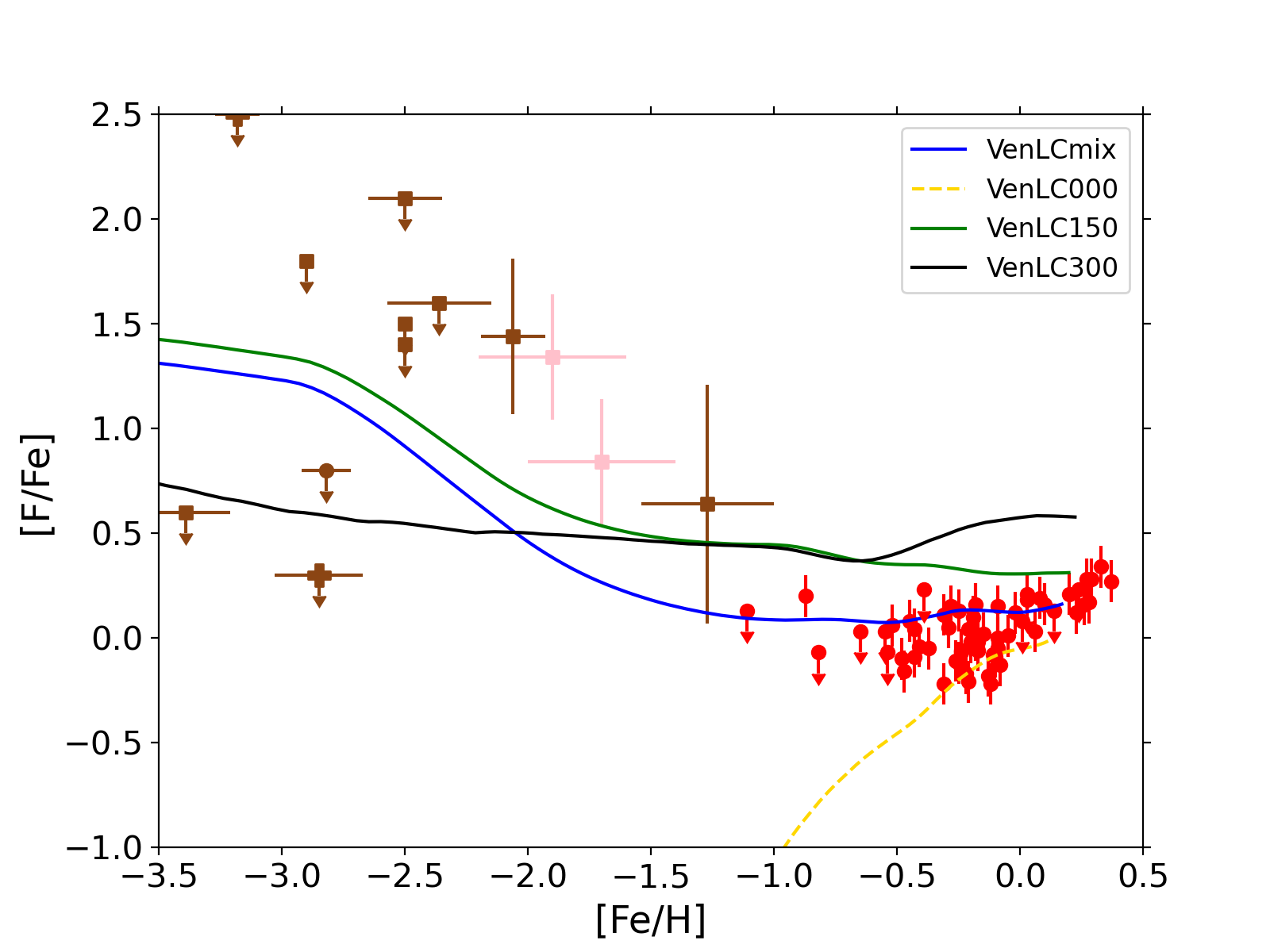}
 \caption{[F/Fe] versus [Fe/H]}
 \label{subfig:Ven_FFe}
\end{subfigure}
\hfill
\begin{subfigure}[t]{0.49\textwidth}
 \includegraphics[width=\textwidth]{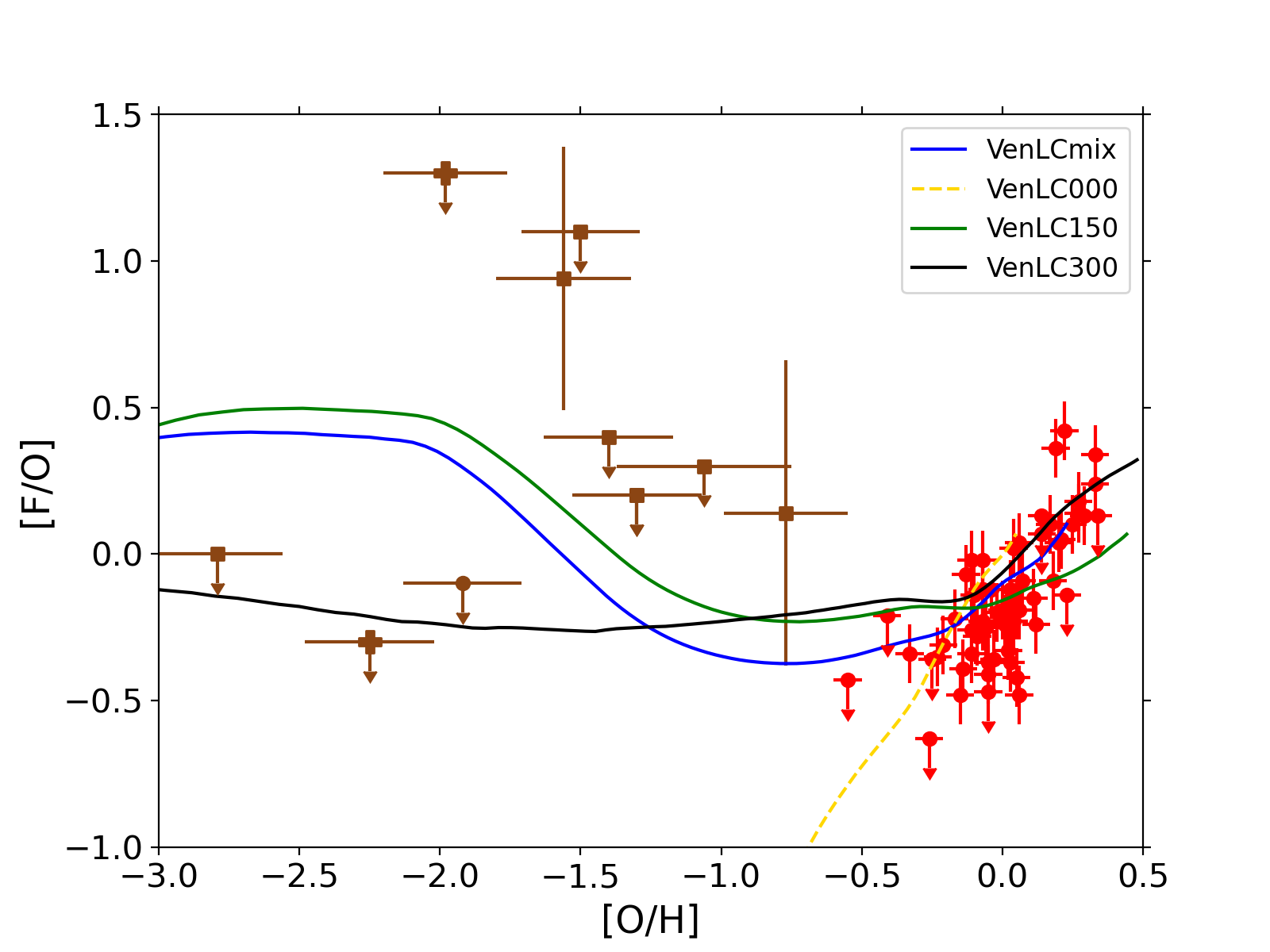}
 \caption{[F/O] versus [O/H]}
 \label{subfig:Ven_FO}
\end{subfigure}
 \caption{Panel a: [F/Fe] versus [Fe/H] for the \citet{Ventura2013, Ventura2014, Ventura2018} AGB yields in combination with each rotational prescription of the \citet{Limongi2018} massive stars, compared with the same observational data as Fig.~\ref{fig:FFe_FeH}. Panel b: same as the Fig.~\ref{subfig:Ven_FFe} but for [F/O] versus [O/H].}
 \label{fig:Ventura_LC}
\end{figure}

The AGB stellar yields of \citet{Ventura2013, Ventura2014, Ventura2018} are explored in more detail in Fig.~\ref{fig:Ventura_LC}, which shows chemical evolution models combining those yields with massive star models with different initial $v_{\text{rot}}$ from \citet{Limongi2018}. Fig.~\ref{subfig:Ven_FFe} shows our results for [F/Fe] versus [Fe/H], whereas Fig.~\ref{subfig:Ven_FO} focuses on [F/O] versus [O/H]. In each figure, the chemical evolution trends for each massive star prescription are similar in shape to the trends we predict when assuming the \textit{FRUITY} AGB yields. However, the final values for models \textit{VenLCmix}, \textit{VenLC000}, \textit{VenLC150} and \textit{VenLC300} for both [F/Fe] and [O/H] are systematically higher than \textit{CLCmix}, \textit{CLC000} \textit{CLC150} and \textit{CLC300}.

The AGB stellar yields of \citet{Ventura2013, Ventura2014, Ventura2018} were the reference set adopted by \citet{Grisoni2020} in their `parallel' chemical evolution model for the Solar Neighbourhood.
However, when comparing our results with those in fig. 1 of \citet{Grisoni2020}, we caution the readers that we assume a different IMF and DTD for SNe Ia. In particular, \citet{Grisoni2020} assumed the IMF of \citet{Kroupa1993}, which hosts much lower numbers of massive stars than the IMF of \citet{Kroupa2001} which we use in our models (see also \citealt{vincenzo2016} for more details); secondly, while \citet{Grisoni2020} assumed the SN Ia single-degenerate model of \citet{Matteucci2001}, here we assume a power-law DTD which is motivated by recent observational surveys \citep[see also \citealt{Wiseman2021} for an observational perspective, and \citealt{Vincenzo2017} for the impact of those two different DTDs on elemental abundance trends]{Maoz2012}.

When we consider the [F/Fe] versus [Fe/H] abundance diagram (Fig.~\ref{subfig:Ven_FFe}), our model with $v_{\text{rot}}=0$ (\textit{VenLC000}) predicts [F/Fe] ratios that are always $\approx$\,0.5\,$\text{dex}$ higher than model \textit{Thin-V000} of \citet{Grisoni2020}. Our models with $v_{\text{rot}}=150$ and $300\,\text{km s}^{-1}$ (\textit{VenLC150} and \textit{VenLC300}, respectively), instead, always lie below models \textit{Thin-V150} and \textit{Thin-V300} of \citet{Grisoni2020} for iron abundances between $-1.5 \lesssim [\text{Fe/H}] \lesssim -0.5$ but then move above them as metallicity increases. It is difficult to compare models with variable $v_{\text{rot}}$ because we follow different prescriptions. We recall that \citet{Grisoni2020} chose the prescription of \citet{Romano2019} with a sharp transition from $v_{\text{rot}}$\,=300\,$\text{km s}^{-1}$ to $v_{\text{rot}}$\,=0\,$\text{km s}^{-1}$ at $\text{[Fe/H]}$\,=\,-1\,$\text{dex}$, whereas our model uses a prescription from \citet{Prantzos2018} which employs a more gradual change to lower rotational velocities as the metallicity increases. The mix of rotational velocities adopted in the present work (\textit{VenLCmix}) follows the observational trends much more closely than \textit{Thin-Vvar} of \citet{Grisoni2020}. 

When we consider the [F/O] versus [O/H] abundance diagram (Fig.~\ref{subfig:Ven_FO}), our models with $v_{\text{rot}}$\,=\,0\,$\text{km s}^{-1}$ always lie above model \textit{Thin-V000} of \citet{Grisoni2020}, being separated by a constant offset of $\approx$\,0.2\,$\text{dex}$. Our model \textit{VenLC150} appears to sit lower than \textit{Thin-V150} of \citet{Grisoni2020} for $[\text{Fe/H}]\lesssim 0\,\text{dex}$. Interestingly, the models with $v_{\text{rot}}$\,=300\,$\text{km s}^{-1}$ follow a very similar shape in both this work and \citet{Grisoni2020}. However, our model always lies below \textit{Thin-V300} of \citet{Grisoni2020}, being separated by an offset of $\approx$\,0.3\,$\text{dex}$. Our model with a rotational mix (\textit{VenLCmix}) can reproduce the observations at high metallicity more closely than the \textit{Thin-Vvar} of \citet{Grisoni2020}. However, we remind the reader once again that each of our works employs a different prescription for rotational mixing.

In summary, this discussion shows how careful we must be when we make chemical evolution models, and it further highlights the uncertainties we have in trying to best model the Milky Way.


\subsection{Fluorine and s-process elements}

The interplay between fluorine and s-process elements has been previously commented on in the literature (e.g. \citealt{Lucatello2011}, \citealt{Abia2011, Abia2015, Abia2019}). Fluorine and s-process elements can be made together both in AGB stars and massive stars, especially when mixing is enhanced by rotation. 

In massive stars, fluorine nucleosynthesis takes place in the helium convective shell via a series of reactions involving $\alpha$-captures and proton captures. s-process elements in massive stars are synthesised via neutron captures, with the neutrons primarily coming from the \el{22}{Ne}\,($\alpha$, n)\,\el{25}{Mg} reaction. \el{22}{Ne} is synthesised from \el{14}{N} produced in the convective H-burning shell and brought into the He-burning core. Once in the core, two convective $\alpha$-captures starting from \el{14}{N} produce \el{22}{Ne}. This process continues into the carbon burning phase (see \citealt{Pignatari2010,Prantzos2018} for more details). In massive star models without rotation, we might expect to see s-process production up to the so-called `first-peak' i.e. Sr, Y, Zr (e.g. \citealt{Limongi2003}). However, \citet{Frischknecht2012} showed that the efficiency of the mixing processes described above can be greatly enhanced within rapidly rotating massive stars, leading to s-process production beyond the first peak.

In AGB stars, both fluorine and s-process elements are made during thermal pulses. Fluorine is made via a series of neutron, proton and $\alpha$ captures that use \el{14}{N} as the seed nucleus. s-process elements are made via neutron captures in the intershell region of the star (e.g. \citealt{Busso1999}). The primary neutron source here is the \el{13}{C}\,($\alpha$, n)\,\el{16}{O} reaction. Given the similar production sites of fluorine and s-process elements, it seems likely that where we find one we would likely find the other. This means there is a potential correlation between fluorine and s-process elements that needs to be explored.

\begin{figure}
    \centering
    \includegraphics[width = \columnwidth]{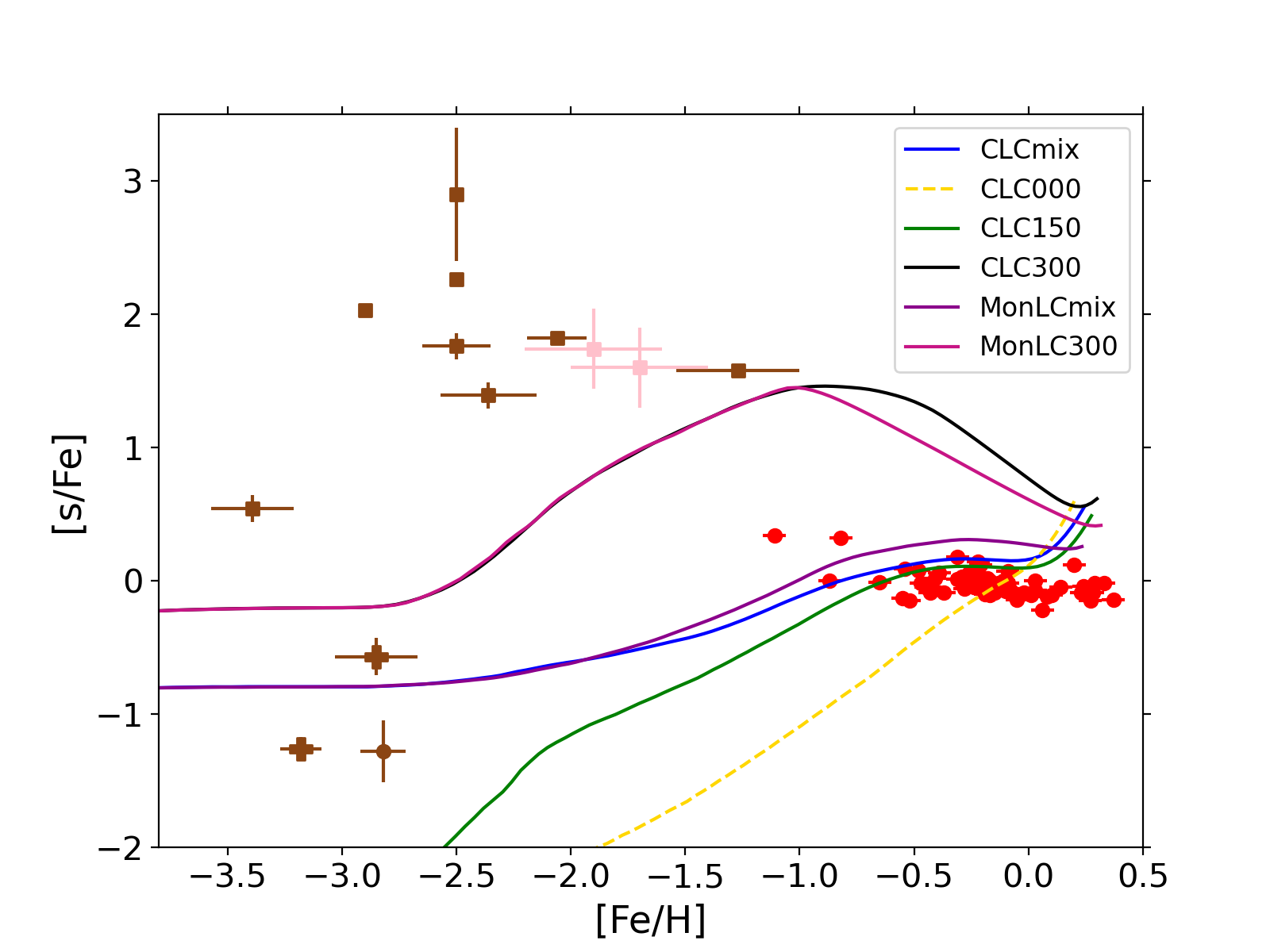}
    \caption{[s/Fe] versus [Fe/H] for models CLCmix, CLC000, CLC150, CLC300, MonLCmix and MonLC300. The observational data is the same as Fig.~\ref{fig:FFe_FeH}.}
    \label{fig:sFe_FeH}
\end{figure}

\begin{figure}
    \centering
    \includegraphics[width = \columnwidth]{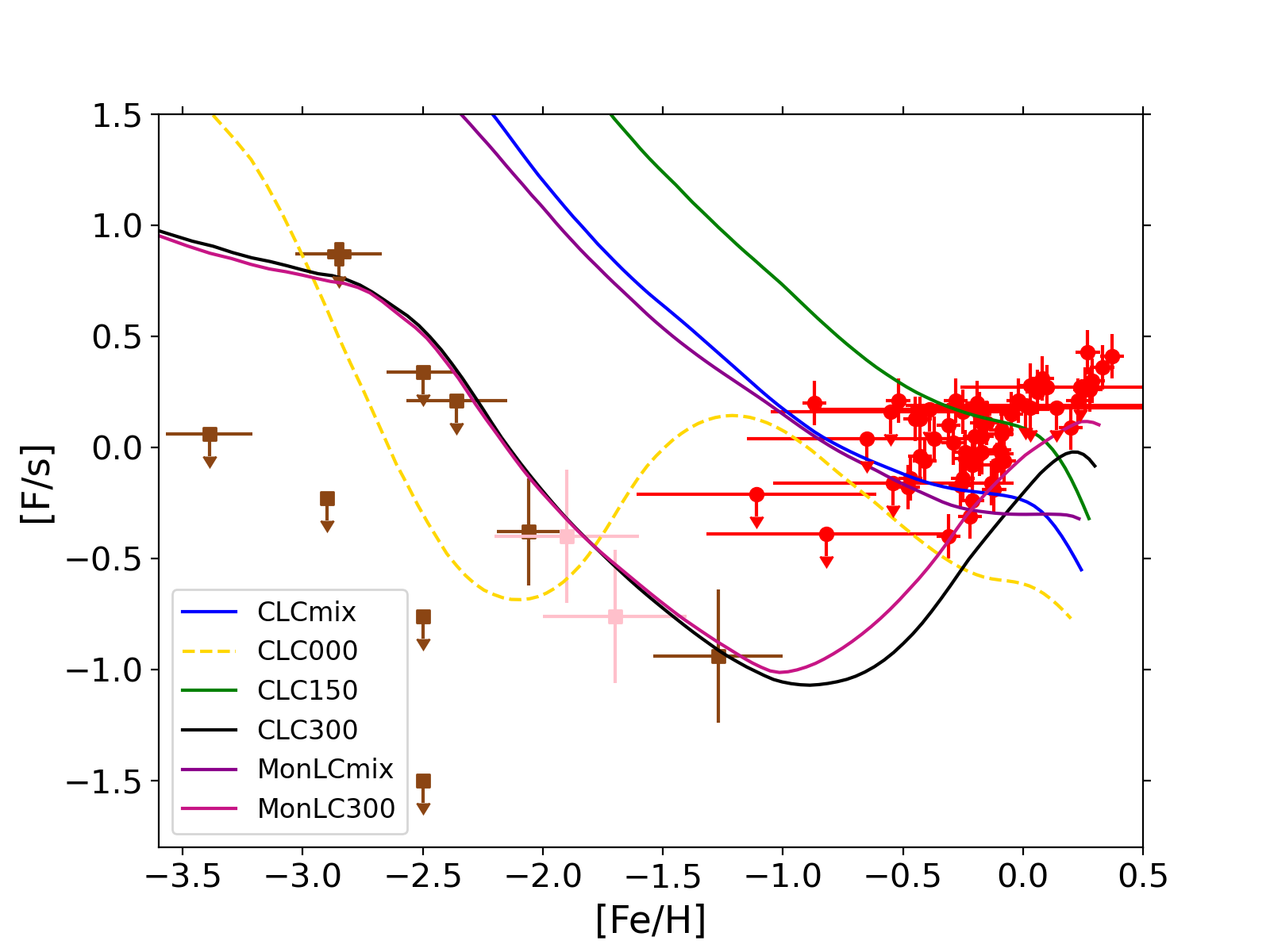}
    \caption{[F/s] versus [Fe/H] for models CLCmix, CLC000, CLC150, CLC300, MonLCmix and MonLC300. The observational data is the same as Fig.~\ref{fig:FFe_FeH}.}
    \label{fig:Fs_FeH}
\end{figure}

Fig.~\ref{fig:sFe_FeH} shows the [s/Fe] versus [Fe/H] abundance trend for the models \textit{CLCmix}, \textit{CLC000}, \textit{CLC150}, \textit{CLC300}, \textit{MonLCmix} and \textit{MonLC300}, that are specified in Table \ref{tab:yields}. The models with the AGB yields of \citet{Ventura2013}, the first set of Monash AGB yields (Mon.1 ), and the massive star yields of \citet{Nomoto2013} are are not shown because they do not include heavy element abundances. In Figs \ref{fig:sFe_FeH} and \ref{fig:Fs_FeH}, `s' denotes the average s-process abundance for each of the models, where [s/Fe] is defined as follows \citep{Abia2002}:
\begin{align*}
    \rm{[s/Fe]\,=\,([Sr/Fe]\,+\,[Y/Fe]\,+[Zr/Fe]\,+\,[Nb/Fe]} \\
    \rm{\,+[Ba/Fe]\,+\,[La/Fe]\,+[Ce/Fe]\,+\,[Pr/Fe])/8}.
\end{align*}
If we focus on the very low-metallicity regime, the only models that can reproduce the high upper limits on [s/Fe] are those which include massive stars with $v_{\text{rot}}=300\,\text{km s}^{-1}$. In the domain -2.5\,$<$\,[Fe/H]\,$<$\,-1 the models \textit{CLC300} and \textit{MonLC300} underestimate the observations. These models also severely overestimate [s/Fe] at high metallicity, disagreeing with the observations of \citet{Ryde2020}. We note that a similar mismatch was also seen by \citet{Vincenzo2021} when comparing their models with the \citet{Limongi2018} rotating massive star yields to the stellar abundance measurements of neutron-capture elements from the second data release of the GALactic Archaeology with HERMES (GALAH) survey \citep{Buder2018}. The rest of the models (those with a minor or absent contribution from stars with $v_{\text{rot}}=300\,\text{km s}^{-1}$) provide a better explanation for the high metallicity observations, with \textit{CLCmix} and \textit{CLC150} reproducing the plateau in the data up to solar metallicity.  

\begin{figure}
    \centering
    \includegraphics[width = \columnwidth]{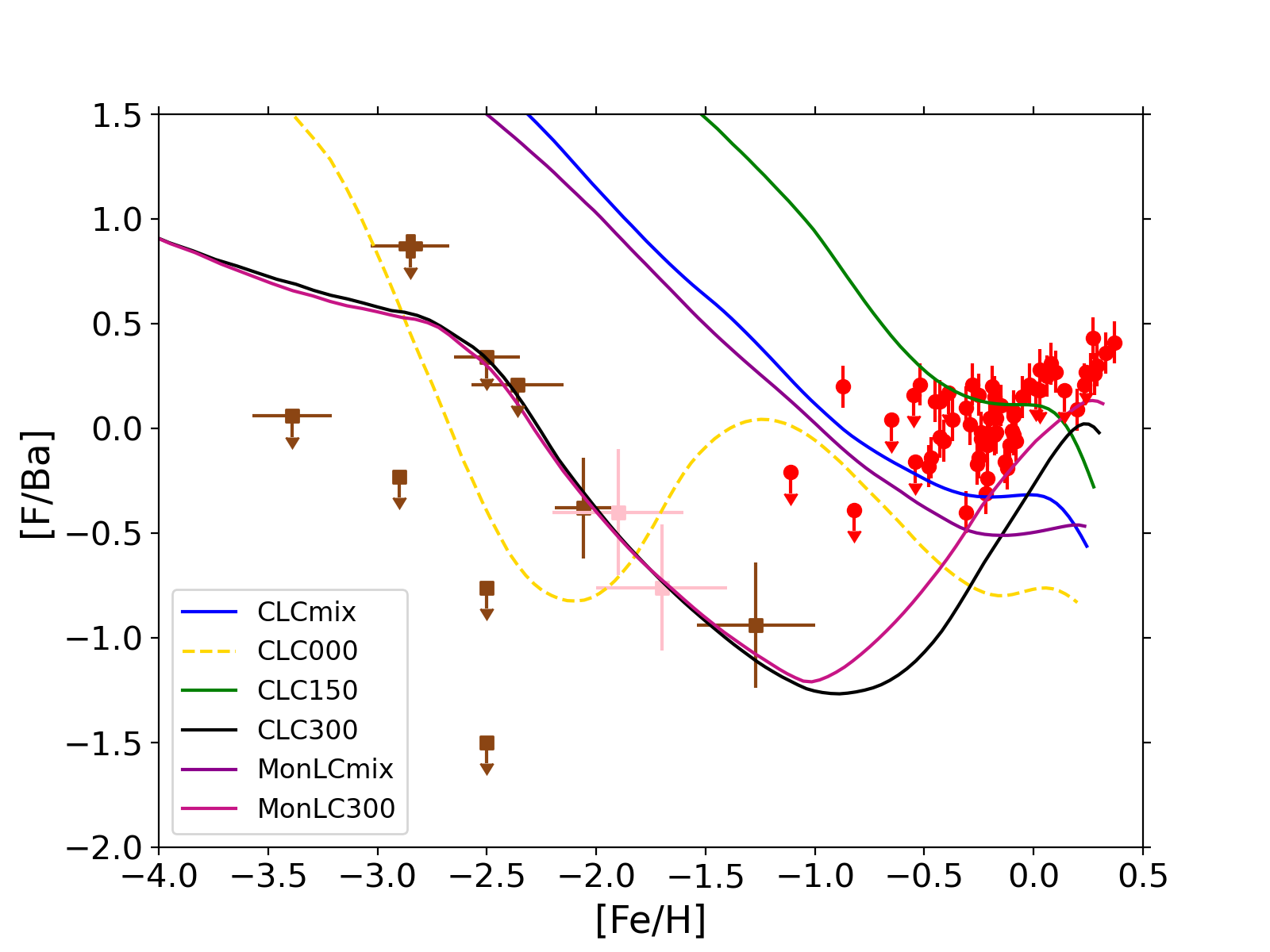}
    \caption{[F/Ba] versus [Fe/H] for models CLCmix, CLC000, CLC150, CLC300, MonLCmix and MonLC300. The observational data is the same as Fig.~\ref{fig:FFe_FeH}.}
    \label{fig:FBa_FeH}
\end{figure}

Fig.~\ref{fig:Fs_FeH} shows [F/s] versus [Fe/H] for the same models as Fig.~\ref{fig:sFe_FeH}. By investigating this ratio we can continue to probe the chemical evolution of fluorine. For comparison, Fig.~\ref{fig:FBa_FeH} shows [F/Ba] versus [Fe/H] for the same set of models. Since there is minimal change in the trajectory of the chemical evolution trends between [F/s] in Fig.~\ref{fig:Fs_FeH} and [F/Ba] in Fig.~\ref{fig:FBa_FeH}, we can safely use the average s-process abundances for comparison with stellar observations by including a variety of s-process elements without loss of important information from tracking elements individually.

In the low metallicity regime (-3.4\,$<$\,[Fe/H]\,$<$\,-2.3), the abundance of F and s-process elements for the CEMP stars in the figure has likely arisen due to accretion of material from an AGB companion (e.g. \citealt{Busso2001}, \citealt{Sneden2008}, \citealt{Lucatello2011}, \citealt{MuraGuzman2020}). Coupled with the fact that most of the observations in this region are upper limits, we cannot use these observations to constrain the GCE models. That being said, it is noteworthy that two scenarios seem to provide similar predictions for [F/s] in Fig.~\ref{fig:Fs_FeH}: (i) AGB + massive stars with $v_{\text{rot}}=0$ (\textit{CLC000}), and (ii) AGB + massive stars with $v_{\text{rot}}=300\,\text{km s}^{-1}$ (\textit{CLC300} and \textit{MonLC300}). These two scenarios are potentially very different. For stars rotating as quickly as $300\,\text{km s}^{-1}$, the fluorine present on the surface will likely have been transported from the interior layers onto the surface due to the strong mixing from rotation. However, internal mixing is not as strong for non-rotating massive stars so there may not be as much fluorine transported from the interior layers to the surface. This could mean some of the surface fluorine is present due to accretion from a companion.

There are two key details in the figures presented in this work that can separate the two potentially different scenarios mentioned above. 
\begin{enumerate}
    \item In Fig.~\ref{fig:Fs_FeH}, the solar and super-solar metallicity observations from \citet{Ryde2020} show an upturn in their [F/s] that is only predicted by the models with $v_{\text{rot}}=300\,\text{km s}^{-1}$.
    \item In Fig.~\ref{fig:FFe_FeH}, model \textit{CLC000} is below all the observations, which means that solely including non-rotating massive stars is not enough to reproduce the observed fluorine abundance pattern.
\end{enumerate}
Overall, this suggests that we need a contribution from rotating massive stars throughout the evolution of the Galaxy in order to reproduce the observations; in particular, Fig.~\ref{fig:Fs_FeH} shows that massive stars with $v_{\text{rot}}=300\,\text{km s}^{-1}$ might play a crucial role in the chemical evolution of fluorine, especially when considering the simultaneous production of s-process elements.

\begin{figure*}
    \centering
    \includegraphics[width=\textwidth]{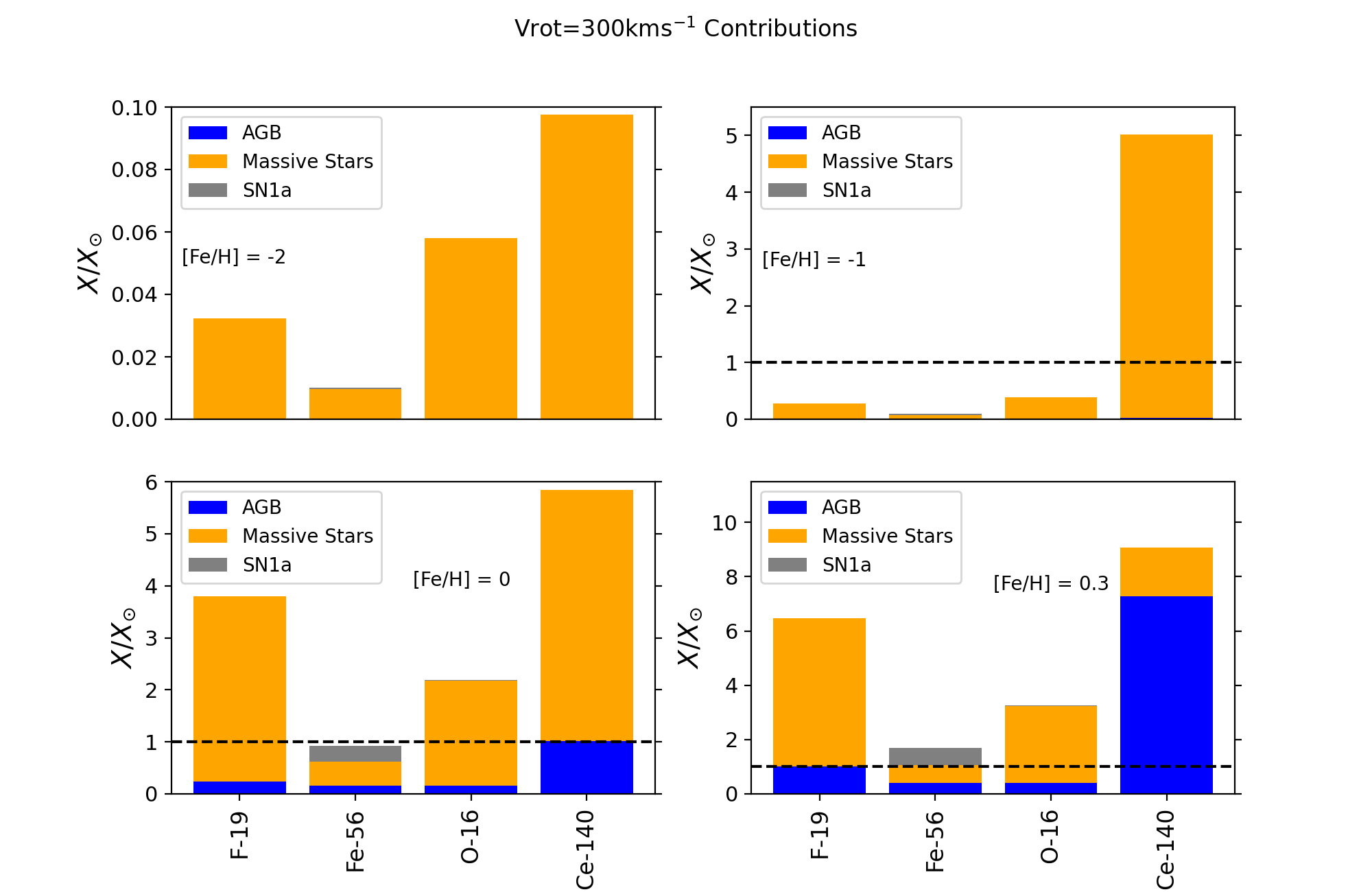}
    \caption{The contribution of each stellar source relative to solar for model CLC300 at metallicities [Fe/H]\,=\,-2 (top left), [Fe/H]\,=\,-1 (top right), [Fe/H]\,=\,0 (bottom left) and [Fe/H]\,=\,0.3 (bottom right - present). The contribution from massive stars is show in orange, AGB stars in blue and SNe 1a in grey.}
    \label{fig:300_cont}
\end{figure*}

\begin{figure*}
    \centering
    \includegraphics[width=\textwidth]{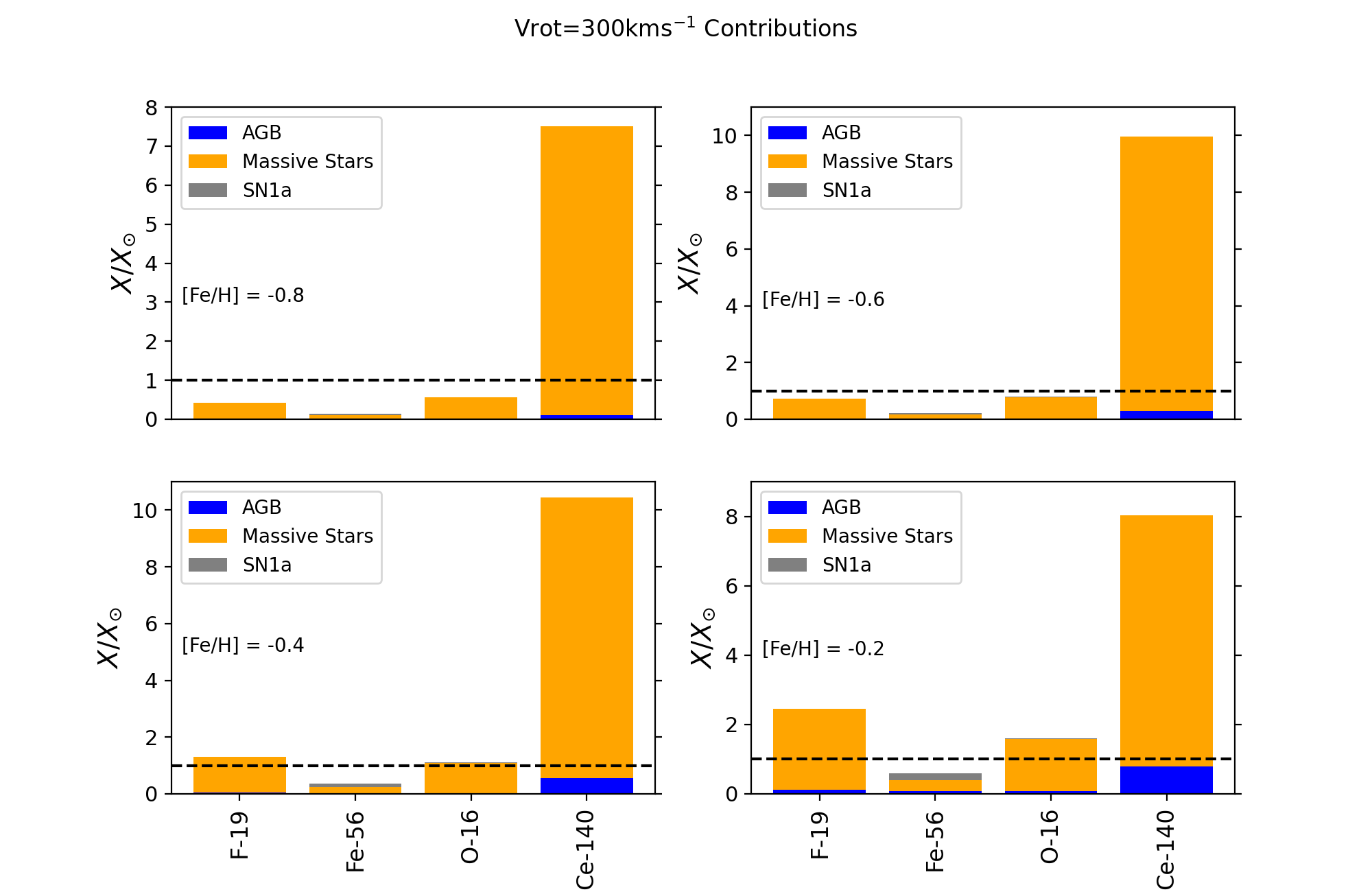}
    \caption{The contribution to each isotope from each stellar source relative to solar for model CLC300 at metallicities [Fe/H]\,=\,-0.8 (top left), [Fe/H]\,=\,-0.6 (top right), [Fe/H]\,=\,-0.4 (bottom left) and [Fe/H]\,=\,-0.2 (bottom right - present). The contribution from massive stars is show in orange, AGB stars in blue and SNe 1a in grey.}
    \label{fig:300_more_detail}
\end{figure*}

Figs. \ref{fig:300_cont} and \ref{fig:300_more_detail} disentangle the contributions from massive stars, AGB stars and Type Ia SNe to \el{19}{F}, \el{56}{Fe}, \el{16}{O}, and \el{140}{Ce} as predicted by the model with $v_{\text{rot}}=300\,\text{km s}^{-1}$ (\textit{CLC300}). Here cerium is used as a proxy for the s-process elements. We can see that the massive star model with $v_{\text{rot}}=300\,\text{km}\,\text{s}^{-1}$ dominates both \el{19}{F} and \el{140}{Ce} even when AGB stars kick in between -1\,$<$\,[Fe/H]\,$<$\,0. Fig.~\ref{fig:300_more_detail} highlights this range in more detail. 

\begin{figure*}
    \centering
    \includegraphics[width=\textwidth]{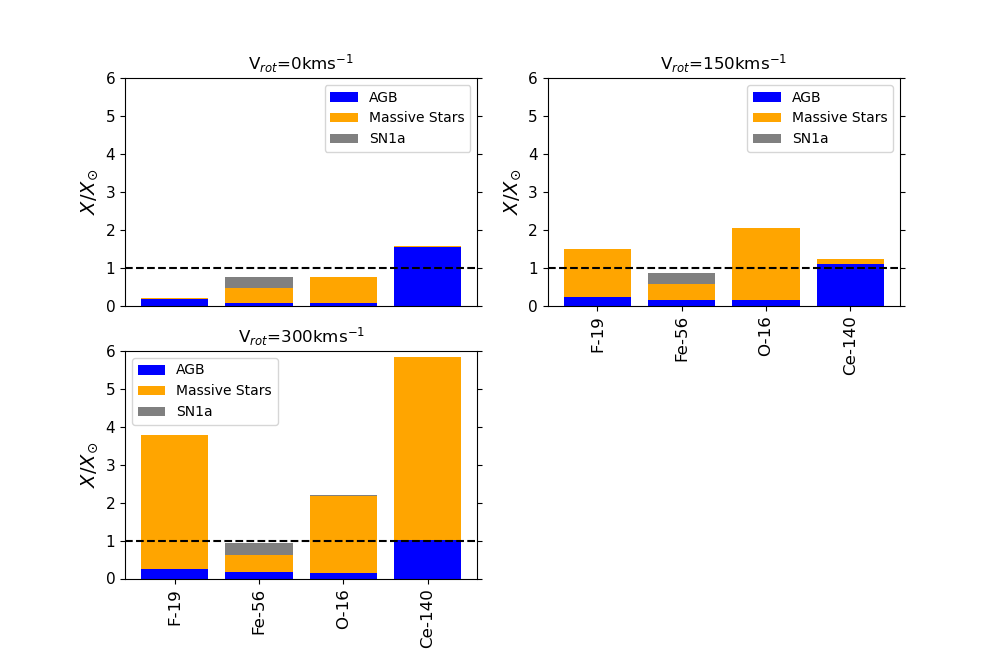}
    \caption{The contribution to each isotope from each stellar source relative to solar at [Fe/H]\,=\,0 for models \textit{CLC000} (top left), \textit{CLC150} (top right) and \textit{CLC300} (bottom). The contribution from massive stars is show in orange, AGB stars in blue and SNe 1a in grey.}
    \label{fig:cont_at_zero}
\end{figure*}

By looking at the predictions of model \textit{CLC300} in Figs. \ref{fig:300_cont} and \ref{fig:300_more_detail}, both \el{19}{F} and \el{140}{Ce} abundances at $[\text{Fe/H}]=0$ are higher than solar by a factor of $\approx 4$ and $\approx 6$, respectively (the black dashed line on each panel corresponds to the solar fluorine and cerium abundances). Therefore, even though models \textit{CLC300} and \textit{MonLC300} are best at reproducing the observational trends of Fig.~\ref{fig:Fs_FeH}, the fluorine and average s-process abundances that they generate at solar metallicity are not physical, suggesting that a mix of massive star models with different $v_{\text{rot}}$ should be assumed. 
The mix of rotational velocity we might expect will be discussed in the following section.

\section{Discussion}\label{sec:discussion}
At low metallicity, most red giants in the sample of \citet{Lucatello2011} are classified as CEMP-s, hence they likely had their surface fluorine abundances altered by binary mass transfer from an AGB companion. In Figs. \ref{fig:FFe_FeH} - \ref{fig:FBa_FeH} we also show CEMP-no stars, whose origin in the Milky Way halo is less clear. Our model predictions at low metallicity can solely be used as a baseline for the average ISM abundances at the point of birth of the stars, before any binary accretion has occurred, providing an empirical constraint on the degree of fluorine-enhancement for AGB stellar models. There is also a larger spread in the observed chemical abundance patterns at [Fe/H]\,<\,-2.5, which indicates a more inhomogeneous ISM at low metallicity, as stars formed out of gas enriched by a smaller number of CCSNe, whereas our models assume that the ISM is well mixed at all times, with the IMF being fully sampled starting from the turn-off mass. An additional source of scatter in the chemical abundances at [Fe/H]\,<\,-2.5, which is not included in our models, might be due to the fact that the Milky Way halo comprises several populations of stars that were born in different substructures and were later accreted by our Galaxy. We also note again that the observations in the metallicity range -3.4\,$<$\,[Fe/H]\,$<$\,-2.3 are upper limits with a lot of dispersion. All this leads to uncertainty in our conclusions at [Fe/H]\,$<$\,-2.

At super-solar metallicity, there is a secondary behaviour of fluorine \citep{Ryde2020}. However, we must be careful about comparing our models to observations at this metallicity for a number of reasons. The first being that we do not have fluorine yields at super-solar metallicity, instead at [Fe/H]\,$>$\,0 the model copies the yields from the final metallicity until the end of the simulation (when the age of the galaxy is 13\,Gyr). Secondly, stars with super-solar metallicity are known to have formed in the inner disk and migrated so their composition is different to that of the local gas (see fig. 10 of \citealt{Vincenzo2020} for an illustration of this). Therefore, the abundances of super-solar metallicity stars cannot be compared with a one-zone model. Though we do not make strong conclusions about the evolution of fluorine above [Fe/H]\,=\,0, these considerations should be kept in mind.

The models using massive stars with initial rotational velocities of $300\,\text{km s}^{-1}$ are the only ones to reproduce both the slight downward trend of [F/s] at low metallicity and upward trend of [F/s] at high metallicities. Therefore, we need a contribution from rapidly rotating massive stars with initial rotational velocities of $300\,\text{km s}^{-1}$ throughout the evolution of the Galaxy in order to match the full abundance pattern. 
Though models \textit{CLC300} and \textit{MonLC300} do not match the full abundance trend of the observations in the [F/Fe] versus [Fe/H] space (Fig.~\ref{fig:FFe_FeH}), there are many considerations to be made including the fact that the low metallicity observations are upper limits so there is a chance that those observations could sit lower than where they are placed, and we expect fewer massive stars rotating that quickly at higher metallicities (see \citealt{Meynet1997}, \citealt{Prantzos2018}). Therefore, we should explore the possibility of a mix of initial rotational velocities, where stars with $v_{\text{rot}}$ in the range $150$-$300\,\text{km s}^{-1}$ contribute throughout the evolution of the Galaxy. \citet{Romano2019} assumed a sharp transition for massive star rotation where massive stars have $v_{\text{rot}}=300\,\text{km s}^{-1}$ for $\text{[Fe/H]}$\,<\,-1 and, suddenly, $v_{\text{rot}}=0$ for $\text{[Fe/H]}\geq -1.$ This strategy is not appropriate for the situation we have here, as a contribution from models with $v_{\text{rot}}$ in the range $150$-$300\,\text{km s}^{-1}$ needs to be assumed even above [Fe/H]\,=\,-1. Given we know that at higher metallicities massive stars should rotate more slowly, perhaps a combination of rotational velocities are present at higher metallicities, much like the approach employed by \citet{Prantzos2018}. 

The mixed-rotation scenario of \citet{Prantzos2018} assumes that rotating massive stars with $v_{\text{rot}}=300\,\text{km s}^{-1}$ cease to contribute to the yields at around [Fe/H]\,$\approx$\,-2, failing to reproduce the observed trend of [F/s] as a function of [Fe/H] (see model \textit{CLCmix} in Fig. \ref{fig:Fs_FeH}). Therefore, a different combination of rotating massive star models needs to be employed, by including a metallicity-dependent contribution from models with $v_{\text{rot}}=150\,\text{km s}^{-1}$ and $300\,\text{km s}^{-1}$ up to solar metallicity. Fig.~\ref{fig:cont_at_zero} shows the contributions of each rotational velocity to the isotopes \el{19}{F}, \el{56}{Fe}, \el{16}{O} and \el{140}{Ce} relative to solar for models \textit{CLC000}, \textit{CLC150} and \textit{CLC300} at $[\text{Fe/H}]$\,=\,0. Model \textit{CLC000} predicts  $X(^{19}\text{F})/X_{\sun}(^{19}\text{F})$\,=\,0.2 and $X(^{140}\text{Ce})/X_{\sun}(^{140}\text{Ce})$\,=1.5\,, model \textit{CLC150} predicts $X(^{19}\text{F})/X_{\sun}(^{19}\text{F})$\,=\,1.4 and $X(^{140}\text{Ce})/X_{\sun}(^{140}\text{Ce})$\,=\,1.2 and model \textit{CLC300} predicts $X(^{19}\text{F})/X_{\sun}(^{19}\text{F})$\,=\,3.7, and $X(^{140}\text{Ce})/X_{\sun}(^{140}\text{Ce})$\,=\,5.8, at [Fe/H]\,=\,0. In order to reproduce the fluorine solar abundance, we need to achieve $X(^{19}\text{F})/X_{\sun}(^{19}\text{F})=1.0$. This can be done with a 45\% contribution from v$_{\rm{rot}}$\,=\,0\,km\,s$^{-1}$, a 50\% contribution from v$_{\rm{rot}}$\,=\,150\,km\,s$^{-1}$ and a 5\% contribution from v$_{\rm{rot}}$\,=\,300\,km\,s$^{-1}$. When employing these contributions, we achieve [F/Fe]\,=\,0.08, [F/O]\,=\,-0.033 and [F/s]\,=\,-0.45. These percentage contributions are our suggestion for a mix of rotational velocities that are successful at reproducing fluorine abundances at solar metallicity. It is difficult to make a suggestion for combinations at other metallicities as we do not have a constraint for the abundances. We must be careful when suggesting a combination of rotational velocities as there are uncertainties in the yields that we must be aware of. Firstly, a change in the implementation of rotation may change the yields of elements affected by rotation. As discussed by \citet{Prantzos2018}, another uncertainty associated with the \citet{Limongi2018} yields in particular is the enhancement of fluorine in the 15\,M\,$_{\rm{\sun}}$ and 20\,M\,$_{\rm{\sun}}$ models with $v_{\text{rot}}$\,=\,150\,kms$^{-1}$. In these models, a smaller He convective shell forms separately to the main He convective shell. When these two shells merge, the base of the new shell is deeper and thus, is exposed to higher temperatures which causes an enhancement in fluorine production. It is pointed out by \citet{Prantzos2018} that is is difficult to know if this scenario is `realistic' given it only affects two of the stellar models. Other uncertainties such as reaction rates and nuclear networks will be discussed later in this work.

It has been proposed that a contribution from novae is needed in order to match the observed behaviour of [F/O] versus [O/H] (e.g. \citealt{Timmes1995}, \citealt{Spitoni2018}). The majority of the models in this work (\textit{CLCmix}, \textit{CLC150}, \textit{CLC300}, \textit{Mon18 LCmix}, and \textit{MonLCmix}) can reproduce the trends of [F/O] versus [O/H] without including any chemical enrichment of fluorine from novae (see Fig.~\ref{fig:FO_OH}). Therefore, it could be argued that we no longer need a contribution from novae to understand the chemical evolution of fluorine. That being said, it is important to understand the fluorine yields we might expect from novae and the consequences that could have on our results. It is unclear from the literature both how frequent the occurrence of novae are and the fluorine yields we might get from them. \citet{Kwash2021} suggests a nova rate of $\approx 30 \,\text{yr}^{-1}$ while \citet{Shafter2017} suggests a nova rate of $\approx 50 \,\text{yr}^{-1}$ and recent results from \citet{Rector2022} suggest a rate of $\approx 40 \,\text{yr}^{-1}$, however this result is for M31 rather than the Milky Way. Both \citet{Spitoni2018} and \citet{Grisoni2020} used the nova yields as predicted by \citet{Jose1998}, who found that fluorine is only significantly synthesised in their $1.35\,\text{M}_{\sun}$ model, with a maximum yield of $5.4\times 10^{-5}\,\text{M}_{\sun}$ and a minimum yield of $9.9\times 10^{-7}\,\text{M}_{\sun}$. This gives a range of potential \el{19}{F} nova production rate that varies between $2.97\times 10^{-5}\,\text{M}_{\sun}\,\text{yr}^{-1}$ and $2.7\times 10^{-3}\,\text{M}_{\sun}\,\text{yr}^{-1}$. The upper bound here is so high due to the significant yield from the $1.35\,\text{M}_{\sun}$ model. This wide range makes the contribution of novae to the galactic fluorine very uncertain. However, we can compare the potential novae yields to the yields we might expect from CCSNe. The CCSNe rate is variable with time in our model with an average rate of $ 0.025\,\text{yr}^{-1}$. The minimum \el{19}{F} yield from the \citet{Limongi2018} massive star yields with $v_{\text{rot}}$\,=\,300\,$\text{km s}^{-1}$ is $1.027\times 10^{-5}\,\text{M}_{\sun}$ and the maximum is $1.025\times 10^{-3}\,\text{M}_{\sun}$. This yields a potential range of \el{19}{F} production rate from CCSNe between $2.57\times 10^{-7}\,\text{M}_{\sun}\,\text{yr}^{-1}$ and $2.56\times 10^{-5}\,\text{M}_{\sun}\,\text{yr}^{-1}$. This range is lower than that of the potential nova yields. However, we must be aware that only the $1.35\,\text{M}_{\sun}$ nova model is enhanced in fluorine, so there is the potential for the range of fluorine yield from novae to be lowered given that the enhancement only occurs at this one particular mass. \citet{Starrfield2020} looked at \el{19}{F} ejecta from novae for a 1.35\,M$_{\rm{\sun}}$ star and found a range of 6.3\,$\times$10$^{-11}$\,M$_{\sun}$ to 1.0\,$\times$\,10$^{-6}$\,M$_{\sun}$, again demonstrating how uncertain fluorine yields from novae can be. Overall, we recognise that novae may indeed contribute to the galactic fluorine, though the yields are highly uncertain and several critical assumptions need to be made to include them in chemical evolution models, however, they are not required to reproduce the observational abundance patterns in this work.

\subsection{Wolf-Rayet stars as a significant source of fluorine?}
When massive stars rotate, they can, even if only for a brief period, enter into a Wolf-Rayet phase. Given that WR winds have been suggested as a dominant contributor to the chemical evolution of fluorine (\citealt{Meynet2000}, \citealt{Renda2004}), it is important to disentangle what portion of the rotating massive star yields comes from WR winds and what portion comes from the CCSN at the end of their evolution.

\citet{Meynet2000} found that WR stars could contribute significantly to the galactic fluorine content by calculating a series of WR yields and incorporating them in a chemical evolution model for the Milky Way. They found that the \el{19}{F} wind yield of a 60\,M$_{\rm{\sun}}$ model could be a factor of $10$-$70$ times higher than the initial stellar content of \el{19}{F}. These fluorine yields were subsequently used in the chemical evolution study of \citet{Renda2004}, who explored three different scenarios for the nucleosynthesis of fluorine by using the chemical evolution code {\tt GEtool} (\citealt{Fenner2003}, \citealt{Gibson2003}). The first scenario explored by \citet{Renda2004} used solely yields from CCSNe, the second CCSNe and WR stars, and the third used CCSNe, WR and AGB stars. \citet{Renda2004} concluded that, while AGB stars dominate fluorine production at low metallicity, WR stars are the dominant source of fluorine at solar and super-solar metallicities (see their fig. 4). In the years since, many more massive star models have been created which include WR yields. This begs the question, do any of these studies find \el{19}{F} yields as high as those found by \citet{Meynet2000}?

\begin{figure}
    \centering
    \includegraphics[width = \columnwidth]{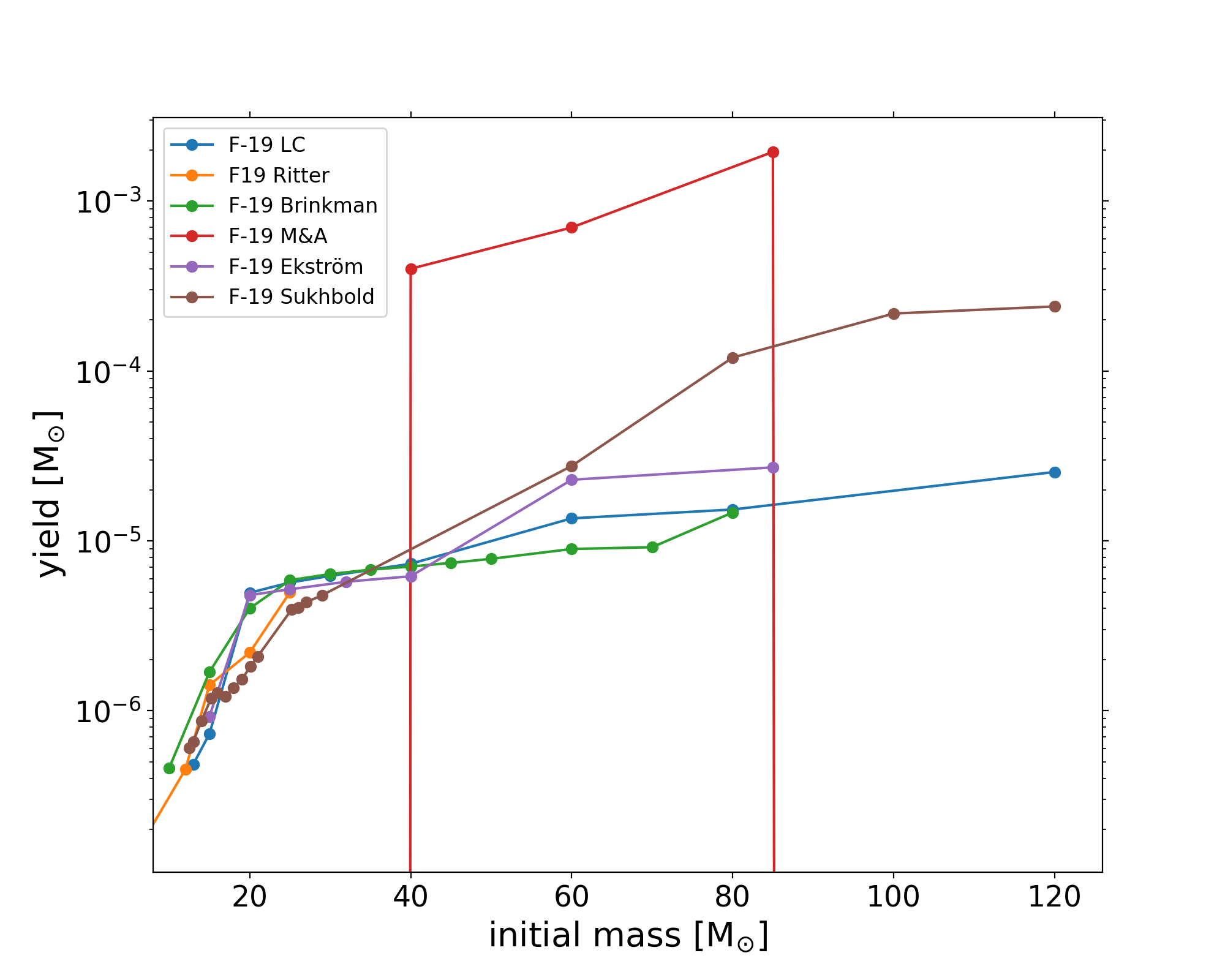}
    \caption{Non-rotating, solar metallicity wind yields from a range of studies over the last two decades. LC = \citet{Limongi2018}, Ritter = \citet{Ritter2018a}, Brinkman = \citet{Brinkman2021, Brinkman2022}, M\&A = \citet{Meynet2000}, Ekstr{\"o}m = \citet{Ekstrom2012}, Sukhbold = \citet{Sukhbold2016}}
    \label{fig:wr_comp}
\end{figure}

Fig.~\ref{fig:wr_comp} shows a comparison of massive star wind yields from a variety of studies over the last couple of decades. The yields that are compared in the figure are from \citet{Limongi2018}, \citet{Ritter2018a}, \citet{Brinkman2021,Brinkman2022}, \citet{Meynet2000}, \citet{Ekstrom2012} and \citet{Sukhbold2016}. Here, we look at non-rotating stars at solar metallicity in order to gain the widest comparison and to be able to compare with the non-rotating yields of \citet{Meynet2000}. We can see that all considered wind yields sit at least $1\,\text{dex}$ below the the \citet{Meynet2000} yields. This suggests that perhaps the \citet{Meynet2000} \el{19}{F} wind yields are unusually high compared to subsequent models. Therefore, there is potential that we may be able to rule out WR stars as a dominant contributor to the galactic fluorine budget.

\begin{figure}
    \centering
    \includegraphics[width=\columnwidth]{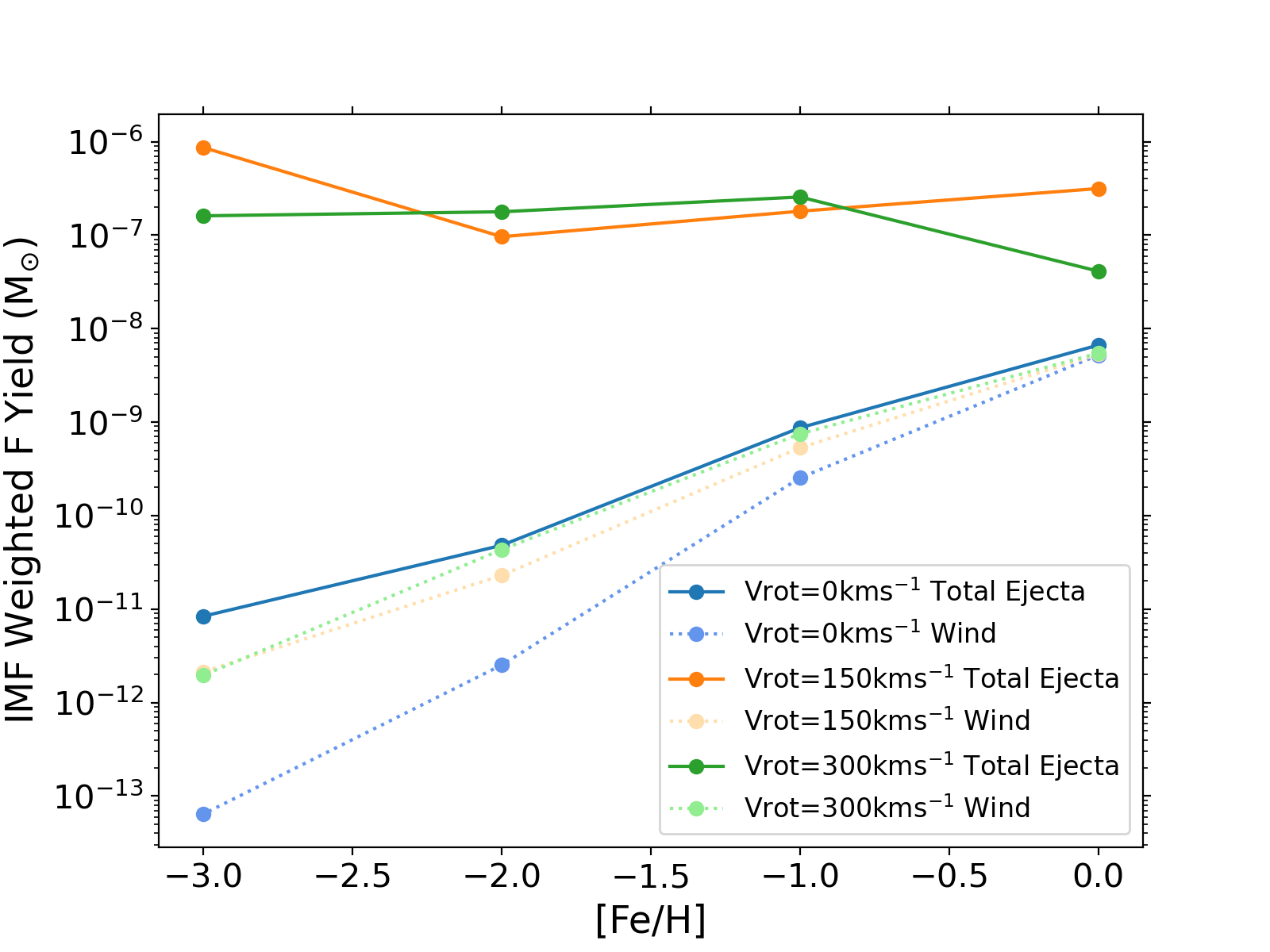}
    \caption{IMF weighted fluorine yield as a function of metallicity ([Fe/H]) for each rotational velocity prescription. The total yields are shown by a full line and darker colour while the wind yields are represented by dotted lines and lighter colours.}
    \label{fig:yield_v_met}
\end{figure}

To investigate this further, Fig.~\ref{fig:yield_v_met} shows the IMF-weighted yield versus metallicity for the \citet{Limongi2018} massive star yields used in this work. Here we see a comparison between the wind yield and the total ejecta for each rotational velocity. At low metallicity, the wind yields sit 4-6\,dex lower than the total ejecta for the rotating models and around 2\,dex lower for the non-rotating model. At higher metallicities, the gap between wind contribution and total ejecta reduces due to enhanced mass loss, with the wind yields being around 2\,dex lower than the total ejecta for the model rotating at 300\,km\,s$^{-1}$ and 1\,dex lower for the model rotating at 150\,km\,s$^{-1}$. For the non-rotating model, the yields are almost identical at high metallicity.

We conclude that we need a large contribution from rapidly rotating massive stars in order to reproduce observations of fluorine in the Milky Way across the whole metallicity range. For $v_{\rm{rot}}$\,=\,300\,km\,s$^{-1}$, wind yields contribute a factor of around 10$^{-2}$ less fluorine at high metallicities ([Fe/H]\,$\geq$\,0) and a factor of around 10$^{-6}$ less fluorine at the lowest metallicity ([Fe/H]\,=\,-3) than the explosive yield, for $v_{\rm{rot}}$\,=\,150\,km\,s$^{-1}$ wind yields contribute a factor of around 10$^{-1}$ less fluorine at high metallicities and a factor of around 10$^{-6}$ less fluorine at the lowest metallicity. We can, therefore, rule out Wolf-Rayet stars as a dominant source of fluorine. Being able to draw such conclusions is vital in untangling the web of possibilities for the origin and chemical evolution of fluorine.

\subsection{Sources of Uncertainty in GCE}
Like for any physical model, it is important to consider that there can be significant uncertainties concerning galactic chemical evolution (GCE) studies (e.g., \citealt{Romano2005, Romano2010}). For example, each choice for the parameters in Table \ref{tab:model_properties} can affect the behaviour of the chemical evolution models. \citet{Cote2016} explored some sources of GCE uncertainty, including: the IMF , DTD and number of SNe Ia, current stellar mass and star formation history. 

We briefly explored the effect of changing both the IMF and the star formation efficiency of the models on our results. We found that:
\begin{enumerate}
    \item using a \citet{Kroupa1993} IMF rather than \citet{Kroupa2001} does not drastically change the results of the chemical evolution trends. Using the \citet{Kroupa1993} IMF produces more fluorine at lower metallicities which can produce a slightly better fit for [F/Fe] versus [Fe/H] trends but provides an overproduction of [F/O] as a function of [O/H] for the models which use the \citet{Limongi2018} yields. However, a better fit to the observations is achieved by the model including the massive star yields of \citet{Nomoto2013}.
    \item using a higher star formation efficiency naturally exhausts the available gas more quickly, and thus does not produce as much fluorine at higher metallicities whereas a lower star formation efficiency sees a late increase in [F/Fe]. However, the shape of the chemical evolution trend is not significantly affected.
\end{enumerate}

Another major source of uncertainty in GCE studies are the yield sets used (see, e.g., \citealt{Gibson1997,Molla2015}). Each author will use a different code for stellar modelling which will in turn use a different reaction network. A reaction network specifies the reactions that will occur in a model and the rates at which such reactions will occur. Different modelling choices made by each author produce a layered effect when it comes to the uncertainty provided by stellar yields in chemical evolution modelling.

To better understand reaction rate uncertainties in the context of this work, we will look at the two reactions that can destroy fluorine: \el{19}{F}\,($\alpha$, p)\,\el{22}{Ne} and \el{19}{F}\,(p, $\alpha$)\,\el{16}{O}.

\begin{enumerate}
    \item The most recent work on \el{19}{F}\,(p, $\alpha$)\,\el{16}{O} was performed by \citet{Zhang2021}. By reanalysing experimental data, they found drastically different \el{19}{F}\,(p, $\alpha$)\,\el{16}{O} rates than those recommended by the Nuclear Astrophysics Compilation of Reaction Rate (NACRE) \citep{Angulo1999}. They found rates larger by factors of 36.4, 2.3 and 1.7 for temperatures 0.01, 0.05 and 0.1\,GK respectively. This increased rate naturally leads to the destruction of \el{19}{F} on a scale larger than previously thought. By performing a network calculation at solar metallicity with their recommended new rate for the reaction, the value of \el{19}{F} decreased by up to one order of magnitude. This reaction was directly measured by \citet{Zhang2021a} using the Jinping Underground Nuclear Astrophysics (JUNA) experimental facility. Though the rate they found was 0.2-1.3 times lower than that of their theoretical prediction \citep{Zhang2021}, it is still significantly higher than the accepted rate of \citet{Spyrou2000}. Therefore, we will still expect a larger depletion of fluorine at solar metallicity using this reaction rate.
    \item The most recent work to study \el{19}{F}\,($\alpha$, p)\,\el{22}{Ne} is \citet{Palmerini2019}, who focused on the role that this reaction takes in AGB stars in particular. They found that during thermal pulses, \el{19}{F} can be easily destroyed by $\alpha$-captures; in particular for a 5\,M$_{\rm{\sun}}$ AGB star \el{19}{F} can be destroyed by a factor of 4.
\end{enumerate}
These new discoveries related to the reactions that destroy fluorine could have implications for this work. If indeed, the destruction of fluorine is more enhanced in AGB stars than previously thought, the chemical evolution of fluorine at higher metallicites could be affected. The point at which AGB stars begins to be significant is model dependent. For model \textit{CLC300}, Fig.~\ref{fig:300_cont} shows us that AGB stars begin to be significant in the production of fluorine around solar metallicity. Therefore, we might expect that the [F/Fe], [F/O] and [F/s] ratios studied in this work decrease from [Fe/H]\,=\,0. Whether these reaction rates will also have a significant impact in the destruction of fluorine in rotating massive stars remains to be seen.

Uncertainties around reaction rates are a large source of uncertainty in stellar modelling and the yields we retrieve from those models. All this must be kept in mind when studying galactic chemical evolution. Especially given how uncertain each source's contribution to the galactic fluorine is, uncertainties around reaction rates add another piece to this complex puzzle.

\section{Conclusions}\label{sec:conclusions}
We have studied the chemical evolution of fluorine in the Milky Way. We have used a range of yield sets to try to understand the dominant contributor to the galactic fluorine budget. In order to do this we compared our chemical evolution models to abundance determinations across a wide range of metallicities. The main conclusions of this work are as follows:
\begin{enumerate}
    \item We investigated many combinations of yields with different prescriptions for the rotation of massive stars. Though we are limited by upper limits and poor statistics in the low metallicity regime, we found that in order to reproduce fluorine abundances across the whole metallicity range (-3.4\,$<$\,[Fe/H]\,$<$\,0.4), we need a contribution from rapidly rotating massive stars with initial rotational velocities as high as 300\,kms$^{-1}$. We agree with the results of \citet{Prantzos2018} and \citet{Grisoni2020} that rotating massive stars play a crucial role in the fluorine production up to solar metallicities. We also suggest a combination of initial rotational velocities which can reproduce solar abundances.
    \item We have investigated the contribution of massive star and WR winds to the galactic fluorine budget. We compared the winds of more recent massive star models to the winds of \citet{Meynet2000} and found that we expect to see significantly less fluorine in wind yields than we did 20 years ago.
    \item From the initial study of wind yields, we then looked at the fluorine yields from the winds of the massive stars used in our chemical evolution models. We found that the wind yield can be up to six times lower than the ejecta from the core collapse. Thus, we have ruled out WR winds as a dominant contributor to the galactic fluorine.
    \item We can rule out novae as an important source of galactic fluorine. Our models can successfully reproduce the observational pattern in [F/O] versus [O/H] space and as thus we do not need a contribution from novae that others required in order to reproduce the pattern.
    \item These conclusions, especially those related to the low metallicity regime, could be made stronger by additional observations of fluorine at low metallicity.
\end{enumerate}

To conclude, our study into the chemical evolution of fluorine in the Milky Way has found that rapidly rotating massive stars are the dominant contributor to fluorine. We still need a contribution from AGB stars from [Fe/H]$\approx$-1. We have now been able to rule out Wolf-Rayet stars and novae as a significant contributor to the chemical evolution of fluorine. 

\section*{Acknowledgements}
We thank the reviewer for their comments which improved the quality of this work. KAW, BKG and MP acknowledge the support of the European Union's Horizon 2020 research and innovation programme (ChETEC-INFRA -- Project no. 101008324) and ongoing access to \tt viper\rm, the University of Hull's High Performance Computing Facility. HEB and MP acknowledge support of the ERC via CoG-2016 RADIOSTAR (Grant Agreement 724560). HEB acknowledges support from the Research Foundation Flanders (FWO) under grant agreement G089422N. AK was supported by the Australian Research Council Centre of Excellence for All Sky Astrophysics in 3 Dimensions (ASTRO 3D), through project number CE170100013. MP acknowledges significant support to NuGrid from STFC (through the University of Hull's Consolidated Grant ST/R000840/1) the National Science Foundation (NSF, USA) under grant No. PHY-1430152 (JINA Center for the Evolution of the Elements), the "Lendulet-2014" Program of the Hungarian Academy of Sciences (Hungary), the ChETEC COST Action (CA16117) supported by the European Cooperation in Science and Technology, and the US IReNA Accelnet network (Grant No. OISE-1927130). KAW would like to thank Maria Lugaro and her group at Konkoly Observatory for their hospitality and Lorenzo Roberti for his useful insights. 


\section*{Data Availability}
The data generated for this article will be shared on reasonable request to the corresponding author.
 



\bibliographystyle{mnras}
\bibliography{fluorine} 








\bsp	
\label{lastpage}
\end{document}